\documentclass[10pt, twocolumn]{aastex631}
\usepackage{amsmath}
\usepackage{mathtools}
\usepackage{graphicx}
\usepackage{bm}

\newcommand{\be}{\begin{equation}}
\newcommand{\ee}{\end{equation}}
\newcommand{\ba}{\begin{eqnarray}}
\newcommand{\ea}{\end{eqnarray}}
\newcommand{\nprod}{\dot{N}_{\gamma \gamma \rightarrow e^+ e^-}}
\newcommand{\bnabla}{\mbox{\boldmath$\nabla$}}
\newcommand{\bOmega}{\mbox{\boldmath$\Omega$}}
\newcommand{\nn}{\mbox{} \nonumber \\ \mbox{}}
\newcommand{\Rmax}{{\cal R}}
\newcommand{\rmin}{r_{\rm min}}
\newcommand{\Rstar} {R_{\rm NS}}

\newcommand{\ttot}{t_{\rm tot}}

\begin{document}

\title{Quantum Plasma Creation Near a Magnetar}

\author{Jonathan Zhang}
\affiliation{Department of Physics, McLennan Physical Labs, Toronto, ON M5S 1A7, Canada}
\email{jzhang@physics.utoronto.ca}

\author{Christopher Thompson}
\affiliation{Canadian Institute for Theoretical Astrophysics, 60 St. George St., Toronto, ON M5S 3H8, Canada}

\begin{abstract}
Magnetars in quiescent states continue to emit hard X-rays 
with a power far exceeding the loss of rotational energy.  It has recently
been noted  that this hard X-ray continuum may bear a direct signature of quantum
electrodynamic (QED) effects in magnetic fields stronger than the
Schwinger field ($B_{\rm Q} = 4.4\times 10^{13}$ G).  Where the current
flowing into the magnetosphere is driven by narrow structures
in the solid crust, the $e^\pm$ pair plasma supporting the current 
relaxes to a collisional and trans-relativistic state.  The decay of a pair
into two photons produces a broad, bremsstrahlung-like spectrum of hard X-rays,
similar to that observed and extending up to $0.5-1$ MeV.  The conversion of two gamma rays to a 
pair is further enhanced by a factor $\sim B/B_{\rm Q}$.  Monte Carlo calculations of
pair creation in a dipole magnetic field are presented.  Non-local particle injection is
found to be strong enough to suppress the high voltage that otherwise would accompany
a weaker, global twist; the hard X-rays are mostly emitted away from the magnetic
poles.  Some of the pairs annihilate in an optically thin surface layer.
The prototypical anomalous X-ray pulsar 1E 2259$+$586, which shows
a hard X-ray continuum but relatively weak torque noise, slow spindown, and no radio
emission, is a Rosetta Stone for understanding the magnetar circuit,
consistent with the picture advanced here.  For a $15-60$ keV luminosity as low
as $10^{34}$ erg s$^{-1}$, the polar flux of sub-relativistic pairs produces an optical 
depth $3-30$ to electron cyclotron scattering in the $1-10$ keV band,
reducing the net X-ray polarization.
\end{abstract}


\section{Introduction} \label{sec:intro}

Magnetars are sources of electromagnetic outbursts that are bright enough to be detectable outside the Milky Way, as 
gamma ray transients and radio bursts \citep{Burns2021,CHIME2020,Bochenek2020}.  In contrast with core-collapse 
driven explosions (supernovae and jet-driven gamma ray bursts), the early stages of outbursts
can be closely monitored in nearby Galactic sources \citep{Kaspi2017}.  

The broadband radiative emissions of magnetars are mostly
powered by the decay of non-potential magnetic fields and supporting $\sim 10^{20}$ A currents.
In contrast with rotation-powered pulsars,
the currents flowing outside the star may approach in strength those 
flowing in the interior \citep{Thompson2002,Lyutikov2006,BT2007,Thompson2008b,Beloborodov2009,Parfrey2012,
Chen2017,Thompson2020,Mahlmann2023}. 

Important clues are
provided by the broad and spectrally hard X-ray continuum that dominates the bolometric output of the
brightest quiescent magnetars above $10-20$ keV (\citealt{Kuiper2006,Gotz2006}; \citealt{Enoto2017},
and references therein).  The X-ray power can reach $\sim 10^3$ times the spindown power,
and therefore cannot be explained by the standard pulsar model.

Magnetar X-ray emission carries a direct imprint of quantum electrodynamic (QED)
processes in magnetic fields stronger than the Schwinger field, as normalized by
the electron rest mass ($B_{\rm Q} = m_e^2c^3/e\hbar = 4.4\times 10^{13}$ G).  In this regime,
all basic interactions of electrons, positrons and photons are 
altered qualitatively:  binary collisions are significantly enhanced
and nonlinear effects become directly testable
\citep{Adler1971, Daugherty1980, Kozlenkov1986, Gonthier2014, Kostenko2018, Kostenko2019}.

Collisional $e^\pm$ plasma in super-QED magnetic fields offers a economical explanation 
for the measured hard X-ray continuum.  The annihilation of trans-relativistic $e^\pm$ pairs via
$e^+ + e^- \rightarrow \gamma + \gamma$ generates a broad, bremsstrahlung-like spectrum 
in magnetic fields stronger than $\sim 4\,B_{\rm Q} \sim 2\times 10^{14}$ G \citep{Kostenko2018}.
The emission spectrum can be modified, most readily around the peak, by reabsorption 
by photon collisions and multiple electron scattering.  These effects are included in the QED
Monte Carlo (MC) simulations of \cite{Thompson2020}.  That work is extended here to consider
the non-local effects of $\gamma$ ray collisions.  

In this approach, we uncover a direct connection between the hard spectral component 
observed in quiescent magnetars
and the birefringent properties of the magnetized vacuum.  Hard X-rays only escape the inner 
magnetosphere in the ordinary mode (O-mode):  the extraordinary mode (E-mode or X-mode)
rapidly splits into two O-mode photons, $E \rightarrow O + O$, above an energy
$\sim 40$ keV \citep{Adler1971,TD1995}.

Understanding the mechanism of hard X-ray emission is an essential step toward
a self-consistent description of magnetar electrodynamics.
There are important implications for X-ray polarization
\citep{Taverna2022,Turolla2023,Heyl2024,Rigoselli2024}, peculiar pulsed radio emission \citep{Camilo2006,Camilo2007,Levin2010},
and the mechanism by which the solid crust of the magnetar transmits magnetic stresses to its exterior.

Even while hard X-ray emission and rapid spindown in magnetars are generally correlated, 
their time profiles in individual sources can be strongly decorrelated:  
torque increases and radio emission will frequently lag peak X-ray flux 
by months to years (e.g. \citealt{Archibald2015}).  
The currents driving the bolometric output of a quiescent magnetar may not in many cases
be concentrated close to the magnetic poles -- an interesting clue as to why the
fast radio burst (FRB) phenomenon appears rare in Galactic magnetars \citep{Petroff2022}.

\subsection{Non-local $\gamma-\gamma$ Pair Creation}

In this paper, we use a MC method to evaluate the global distribution of pairs
created by $\gamma + \gamma$ collisions.  The pair creation rate is
sensitive to details of the source spectrum around photon energy $m_ec^2$.
The quiescent hard X-ray continuum is observed to rise
in energy flux up to $\sim 100$ keV \citep{Gotz2006,denHartog2008a,denHartog2008b,Enoto2017}.
There is no sign of a cutoff below 100 keV, 
although earlier, non-contemporaneous COMPTEL measurements are suggestive of
a cutoff above 1 MeV in three sources \citep{Kuiper2006}.
For comparison, the spectrum emitted by a trans-relativistic and collisional $e^\pm$ plasma
is predicted to extend to $\sim 0.5-1$ MeV, but then cut off hyper-exponentially above $\sim m_ec^2$
\citep{Thompson2020}.   

In super-QED magnetic fields, the collision of gamma-rays is significantly enhanced:
the cross section increases by a factor $\sim B/B_{\rm Q}$ compared with vacuum
and the kinematic constraints on photon collisions are also relaxed \citep{Kozlenkov1986, Kostenko2018}.  The measured X-ray output at $\sim 50$ keV, extrapolated upward in energy, implies
a total $e^\pm$ annihilation luminosity reaching $\sim 10^{35}$ or $10^{36}$ erg s$^{-1}$
and a significant optical depth to photon collisions.  

\subsection{Self-consistent Circuit State}

More than one self-consistent state is available to the 
$e^\pm$ plasma supplying the current, potentially leading to large variations in voltage and temporal 
behavior.  When the current density is not too high throughout the magnetosphere,
a double layer structure forms. Electron-positron pairs are self-consistently generated via
scattering of keV surface photons at the electron Landau resonance, the embedded charges
reaching a Lorentz factor $\sim 10^{2-3}$ \citep{BT2007}.

More recently, it has been recognized that stronger (and more localized)
currents can relax to a collisional and trans-relativistic plasma state \citep{Thompson2020}.  
The $e^\pm$ interact repeatedly through the single-
and two-photon annihilation channels. Energy and annihilation equilibrium are reached
when the particle density is larger by a factor $\sim 15$ than the minimum needed to 
support the background magnetic shear.

These distinct circuit configurations are not fully compatible, and 
have differing implications for the source of
the hard X-ray continuum.  The magnetic twist could be concentrated around
the magnetic poles
and supported by outward streaming $e^+$ and $e^-$ \citep{Beloborodov2009,Beloborodov2013a}.  By contrast,
denser plasma structures 
(``twist zones'') that connect to crustal shear bands would be distributed more broadly
over the magnetar surface \citep{Thompson2017}.  We consider the
complementary case where the dominant dissipative structures are anchored
at magnetic latitudes around $45^\circ$.

The gamma rays emitted from collisional plasma will reach much 
(but not necessarily all) of the magnetosphere.  As we show here, the pairs produced non-locally
by secondary gamma ray collisions can easily short out the high-voltage, double-layer structure.

Weaker currents flowing into the outer magnetosphere can, in particular, be sustained by this injection 
of pairs.  The polar region of a magnetar is of interest for several reasons. 
During limited phases of activity, this zone appears to support enhanced magnetic twist that
drives large variations in spindown torque and pulsed radio emission \citep{Archibald2015}, 
and may also on occasion emit fast radio bursts \citep{Zhang2023}.  Scattering of $1-10$ keV
photons at the electron cyclotron resonance also has a significant effect on X-ray pulse
profiles \citep{Fernandez2007,Nobili2008}.

\subsection{1E 2259$+$586 as a Test Case}

The prototypical anomalous X-ray pulsar 1E 2259$+$586 \citep{Fahlman1981} offers some 
relatively clean tests of magnetar electrodynamics.  Its X-ray spectrum should be less
modified by repeated scattering than in other magnetars with stronger external 
currents and higher luminosities above 100 keV.  The polar dipole field inferred from spindown
is also weak enough (about $1.2\times 10^{14}$ G) to imply measurable deviations from the
bremsstrahlung-like spectrum that is expected from $e^\pm$ annihilation in magnetic fields 
approaching $10^{15}$ G. 

In fact, 1E 2259$+$586 has one of the hardest $20-50$ keV spectra detected in quiescent magnetars 
\citep{Vogel2014}, consistent with pair annihilation in a magnetic field near $B_{\rm Q}$
(see Figure 3 of \citealt{Thompson2020}).
It is correspondingly dimmer at 20 keV than more burst-active magnetars
with stronger magnetic fields and flatter high-energy spectra (the Soft Gamma Repeaters or SGRs).

The hard X-ray flux emitted by 1E 2259$+$586 (and extrapolated to $\sim 0.5-1$ MeV)
implies a modest optical depth to gamma ray collisions in the bulk of the magnetosphere.
A first-order MC analysis therefore can be carried out by neglecting the
effects of multiple scattering, which are left to future work.

At the same time, 1E 2259$+$586 has a relatively 
quiet spindown history and a characteristic age much greater than the age of the surrounding supernova
remnant CTB 109 \citep{Dib2014}.    Pulsed radio emission has
never been detected (e.g. \citealt{Coe1994}).
1E 2259$+$586 therefore has an ensemble of properties consistent with the proposition that
the source of the hard X-ray continuum can be separated spatially in a magnetar from
the polar magnetic field bundle that controls spindown torque and radio emission -- in contrast
with the approach described by \cite{Fernandez2007}, \cite{Baring2007} and \cite{Beloborodov2013a}.

Finally, 1E 2259$+$586 shows persistent infrared emission $\sim 10^2$ times brighter than is expected 
from the Rayleigh Jeans tail of the surface 0.4 keV blackbody \citep{Hulleman2001}.
Strong coherent plasma emission in the optical-IR band, which is observed more frequently
than pulsed radio emission, is also a natural consequence of
a collisional plasma state with high density \citep{Eichler2002,Thompson2020}.

\subsection{Plan of the Paper}

Section \ref{sec:QED} reviews the interactions of electrons, positrons and photons in super-QED 
magnetic fields, with a focus on enhanced rates of pair creation and the formation of a collisional 
plasma state.  The magnetospheric configurations that are adopted as backgrounds for our MC
calculations of $\gamma-\gamma$ pair creation are described in Section
\ref{sec:Bfield}, and the gravitational lensing of gamma rays 
in Section \ref{sec:trajectory}.  Our MC method is detailed in Section \ref{sec:mc},
and results are presented in Section \ref{sec:results}.   
We summarize the broader implications of our calculations for the torque behavior, 
electromagnetic spectra, and pulse structure of magnetars in Section \ref{sec:summary}.

Throughout this paper, we use standard notation for electron rest mass $m_e$, magnitude of the 
electron charge $e$, speed of light $c$,
reduced Planck constant $\hbar$, and Newton's constant $G$.  The numerical value of quantity $X$
is normalized as $X = X_n\times 10^n$ in cgs units.

\section{QED Processes and the Magnetar Plasma State}\label{sec:QED}

The plasma state outside an active magnetar is sensitive to details of the interactions between 
electrons, positrons and photons in a super-QED magnetic field.
Here, we describe the enhancement of (i) electron-positron collisions in 
zones of high current density and (ii) gamma ray collisions throughout the rest of the 
magnetosphere.  We then show how compensating losses of pairs 
determine the equilibrium pair density.

\subsection{Enhanced Resistivity from Pair Annihilation}

Dissipation outside a neutron star of radius $R_{\rm NS} \sim 10$ km 
at a rate $L_X \sim 10^{35}-10^{36}$ erg s$^{-1}$ would not ordinarily sustain optically thick 
pair plasma: the radiative compactness defined in terms of the non-magnetic Thomson cross section is
$\sigma_{\rm T}L_X/4\pi m_ec^3\Rstar \lesssim 1$.  However, in a super-QED magnetic field, 
two quantum effects allow a pair plasma to settle into a collisonal and trans-relativistic state.

First, as in radio pulsars, the annihilation of a pair into a single photon,
$e^ + + e^- \rightarrow \gamma$, is allowed by the conservation of canonical momentum ${\bm p} \pm 
e{\bm A}/c$, where ${\bm A}$ is the vector potential associated with the stellar magnetic field.  This 
process is resonant, with a cross section exceeding $\sigma_{\rm T}$ by a factor
$\sim (\alpha_{\rm em} B/B_{\rm Q})^{-1}$, where $\alpha_{\rm em}
\simeq 1/137$ is the fine structure constant \citep{Wunner1979}:
\be\label{eq:sigma1gamma}
\sigma_{\rm ann,1\gamma} = 
{3\pi\over 4\beta\gamma^2\alpha_{\rm em}(B/B_{\rm Q})}\sigma_{\rm T}.
\ee
Here $\beta c$ and $\gamma$ are the center-of-momentum (COM) speed and Lorentz factor
of the $e^\pm$.  The created photon will almost immediately convert back to a pair, over
a microscopic distance.  Half the time, the momenta of the 
regenerated electron and positron are reversed, producing a backscattering.  

The second effect enhancing the pair density involves a suppression of
radiative cooling.  Two photon annihilation of the
pairs, $e^+ + e^- \rightarrow \gamma + \gamma$,
is suppressed by a factor $\sim (B/B_{\rm Q})^{-1}$.  This
means that the equilibrium particle density needed to balance
ohmic heating by backscattering is larger, by a factor 
$f_n \equiv (n_{e^+} + n_{e^-})/n_{\rm crit} \sim 15$, than the minimum density 
$n_{\rm crit} = |\bnabla\times{\bm B}|/4\pi\beta e$ that will support a static magnetic twist 
\citep{Thompson2020}.  In addition, in super-QED magnetic fields, most of the energy released in 
two-photon annihilation is carried by a single photon.  This energetic photon
frequently reconverts to a pair, so that only a fraction of the total
energy is lost to an ``annihilation bremsstrahlung'' photon.

A third effect reducing energy loss by trans-relativistic pairs 
is the appearance of a strong $u$-channel resonance
in the cross section for $\gamma-e^\pm$ scattering \citep{Kostenko2018}. This resonance
is encountered when the
photon energy approaches the threshold for single-photon pair
conversion, $\hbar\omega \lesssim \hbar\omega_{1\gamma} \equiv 2m_ec^2/\sin\theta_{kB}$.
(Here, $\theta_{kB} = \cos^{-1}(\hat k\cdot\hat B)$ is the angle of propagation of
a photon of wavevector ${\bm k} = k\hat k$ with respect to the local magnetic field ${\bm B}$.)  

The net effect is that collisional pair plasma can be sustained in a quiescent magnetar
in narrow flux elements sustaining strong magnetic shear.
Given a shear length $\ell_B = B/|\bnabla\times{\bm B}|$ and 
pair density $n_p = n_{e^+} = n_{e^-}$, the normalized optical depth
for $e^+$ to backscatter off $e^-$ along a length $\ell_\parallel$ of a magnetic arcade is
\ba
{\sigma_{\rm ann,1\gamma}\over 2}n_p\,\ell_\parallel &\;\sim\;&
{\pi\over 8\beta\gamma^2} \left({2n_p\over n_{\rm crit}}\right) {\ell_\parallel\over \ell_B}\nn
 &\;=\;& 6\,\left({f_n\over 15}\right){\ell_\parallel\over \ell_B}
\qquad (\beta \sim 0.6).
 \ea

Small-scale braiding of the magnetic field, with an
amplitude $\delta B < B$, wavenumber $k_\perp \gg \Rstar^{-1}$ and twist density 
$k_\perp \delta B \gg B/\Rstar$, will further enhance the current density and collision rate.  
In this way, the appearance
of collisional plasma outside a quiescent magnetar is diagnostic of the yielding motions
that are driven by Maxwell stresses operating below the stellar surface.
 
Magnetar outbursts
point to crustal dissipation with the required properties: 
fault-like zones of sub-kilometer width
that locally shear the magnetic field.  
Most bright SGR bursts have a duration $\sim 0.1$ s,
comparable to the time for elastic stresses to propagate around
the star \citep{Gogus2001};  they are often followed by
decaying hotspot emission covering a few percent or less of the
magnetar surface \citep{Kaspi2017}.  Pseudofaults with
this property are observed in a global elastic-plastic-thermal
model of the thin\footnote{The thickness of the NS crust weighted
by the depth-dependent shear modulus is typically $\sim 0.3$ km.} crustal shell \citep{Thompson2017}.

Details of the transition from a double layer plasma structure at low current density to a collisional state at high current density remain to be investigated.  Such a transition is expected because (i) the double layer structure is much more dissipative at high current density; (ii) the photon flux resulting from particle bomardment of the magnetar surface by relativistic $e^\pm$ will greatly enhance non-resonant photon-electron drag, leading to runaway growth in photon energy density when $n_{\rm crit}\sigma_{\rm T}R_{\rm NS} > 1$; and (iii) a magnetar may produce collisional plasma during a bright, transient outburst, which is accompanied by hysteresis in the collision rate as the luminosity drops.

   \begin{figure}
       \includegraphics[width = \linewidth]{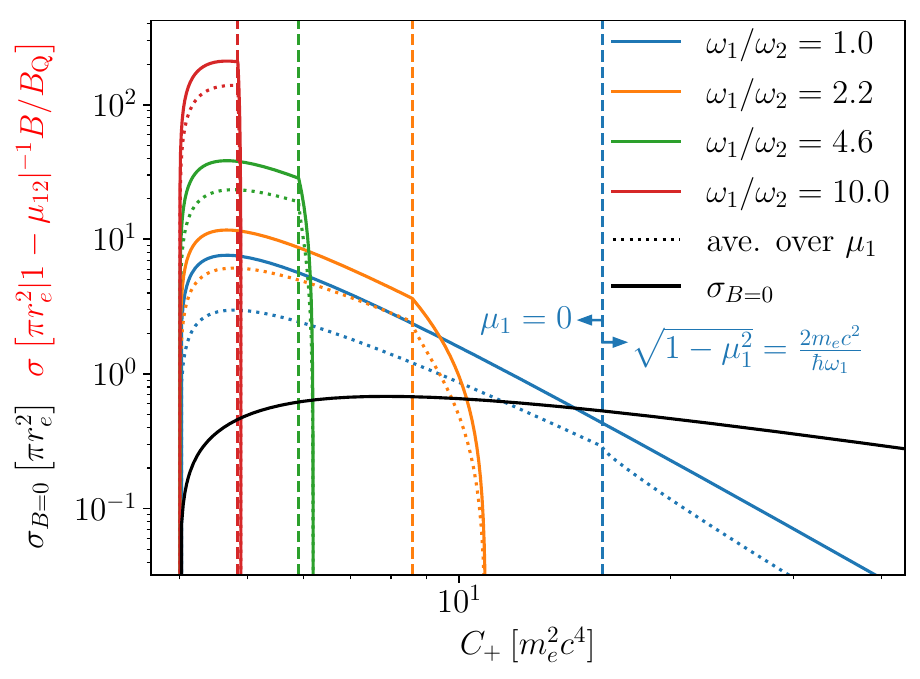}
        \caption{Cross section for $\gamma + \gamma \rightarrow e^+ + e^-$.  Colored lines: 
        the regime $B \gtrsim 4B_{\rm Q}$, plotted as a function of 
        $C_+ = (\hbar\omega_1 + \hbar\omega_2)^2 - (\mu_1 \hbar\omega_1 + \mu_2 \hbar\omega_2)^2$, 
        for various ratios $\omega_1/\omega_2$ of photon energies. Solid lines: cross section at the lowest value of $\mu_1 = \hat k_1\cdot\hat B$
        consistent with $\omega_1 < \omega_{1\gamma} = 2m_ec^2/\hbar(1-\mu_1^2)^{1/2}$; this is $\mu_1 = 0$ to 
        the left of the vertical dashed line, and $\mu_1 > 0$ to the right.  Dotted line: 
        average over all $\mu_1$ for which $\omega_1 < \omega_{1\gamma}$.  Black curve:
        cross section in vanishing magnetic field, now with $C_+ = 2\hbar\omega_1\hbar\omega_2(1- \mu_{12})$.}
        \label{fig:cross_section}
    \end{figure}

\subsection{Photon Interactions and Enhanced Collisions}

Because we are considering dissipative zones with a limited scattering depth ($\lesssim 3-10)$,
our focus is on the emission, propagation, and absorption of the ordinary polarization mode (O-mode).  
The effect of the background magnetic field 
on photon-electron interactions can be understood semi-classically in terms of a suppression of 
${\bm E}\times{\bm B}$ drift of the electron.  The O-mode has electric vector 
partly overlapping the background magnetic field:  
$\delta{\bm E} \propto \hat k\times(\hat k\times {\bm B})$ when the photon propagates
in direction $\hat k$.  The interactions of the E-mode are suppressed by factors
$\sim (\omega m_ec/eB)^2$ at frequency $\omega$ \citep{Harding2006}.  
The cross sections calculated by \cite{Kostenko2018,Kostenko2019} in super-QED magnetic fields
are restricted to the O-mode.

Fundamental to our considerations is the inability of O-mode X-rays
to split, $\gamma \rightarrow \gamma+\gamma$, in the energy range 
$\hbar\omega \gtrsim 40$ keV where the E-mode is rapidly absorbed \citep{Adler1971}.  
The strongly magnetized zone around pulsars or magnetars has birefingent properties
deriving from the nonlinearity in Maxwell's equations that is imparted by vacuum polarization.
The correction to the O-mode index of refraction $kc/\omega$ 
from vacuum polarization is of order $\alpha_{\rm em}B/B_{\rm Q}$
in super-QED magnetic fields, whereas the E-mode only receives a correction
of order $\alpha_{\rm em}$.  This means that energy and momentum can be conserved
in the splitting of a photon into two obliquely propagating photons only when
at least one daughter photon has the larger index of refraction.  The strong break
seen around $10-20$ keV in many quiescent magnetar spectra may be directly related to this QED effect,
if the lower-energy keV surface emission is partly carried by the E-mode.

Let us consider the ensemble of O-mode photons that are below the threshold
for direct conversion to an electron-positron pair, $\omega < \omega_{1\gamma}$.
(This energy depends on the local angle of propagation
$\theta_{kB}$ with respect to ${\bm B}$, meaning that even if
$\omega < \omega_{1\gamma}$ at emission, one must repeatedly check for absorption along
a ray trajectory.)  The volumetric rate of two-photon pair creation is given in terms 
of the cross section $\sigma$ by
    \ba \label{eq:ndotgg}
        \dot{n}_{\gamma \gamma \rightarrow e^+ e^-} &=& \int d\hat k_1 d\hat k_2 d\omega_1 d\omega_2
        \biggl[\frac{d^2n_{\gamma 1}}{d\hat k_1 d\omega_1}\frac{d^2 n_{\gamma 2}}{d\hat k_2 d\omega_2} \nn  
        && \times |1- \mu_{12}| c \, \sigma(\omega_1, \omega_2, \hat k_1, \hat k_2, B)\biggr].
    \ea 
Here, $n_\gamma$ is the photon density, which must be evaluated at each position over a range of
momenta of photons 1 and 2, and $\mu_{12} = \hat k_1\cdot\hat k_2$.

The kinematic constraints on $\gamma + \gamma \rightarrow e^+ + e^-$ are relaxed in a 
super-QED magnetic field. 
In vacuum, the energies of the two photons must pass the threshold
    \be
        C_{+, B=0} \equiv 2 \hbar\omega_1\,\hbar\omega_2 (1 - \mu_{12}) > 4 m_e^2 c^4.
   \ee 
A soft photon of energy $\hbar \omega_s \ll m_e c^2$ may convert to a pair by
colliding with a much harder photon of energy $\hbar \omega \gtrsim 
(m_ec^2)/\hbar\omega_s \gg m_e c^2$.
When $B \gtrsim  B_{\rm Q}$, the same soft photon can 
collide with a photon that individually is not much below the threshold energy
$\hbar\omega_{1\gamma} \sim 2m_ec^2$ in the COM frame.
Pair conversion only requires
    \be \label{eq:threshold}
        C_+ \equiv (\hbar\omega_1 + \hbar\omega_2)^2 - (\mu_1 \hbar \omega_1 + \mu_2 \hbar \omega_2)^2 > 4 m_e^2 c^4, 
        \ee 
where $\mu_i = \hat{k_i} \cdot \hat{B}$.  Note also that two colinear photons
can now convert to a pair, as a result of the breakdown in the conservation of kinetic momentum
perpendicular to ${\bm B}$.

The cross section for $\gamma + \gamma \rightarrow e^+ + e^-$ in magnetic fields stronger than
$\sim 4B_{\rm Q}$ can be written
for arbitrary photon frequencies and direction cosines (not satisfying the COM
condition $\omega_1\mu_1 = -\omega_2\mu_2$) as (\citealt{Kostenko2018}, see Equation 7 of Erratum),
\ba\label{eq:sigmagg}
  \sigma_{\gamma\gamma}[B \gg B_{\rm Q}] = 
  \frac{32\pi r_e^2}{|1 - \mu_{12}|}\frac{B}{B_{\rm Q}}
  \frac{\sqrt{C_+ - 4m_e^2 c^4}}{C_+^{1/2} (\hbar^2\omega_1 \omega_2)^3}\nn
  \times\frac{C_+^2 (1 - \mu_1^2) (1-\mu_2^2)(m_e c^{2})^{6}}
  {[C_+(1-\mu_1^2)(1-\mu_2^2) + 4m_e^2c^4(\mu_1 - \mu_2)^2]^2}.\nn
\ea
One sees from Figure \ref{fig:cross_section} that collisions between soft and 
hard photons have a higher cross section than collisions between photons with 
comparable energy. 

The enhancement in $\sigma_{\gamma\gamma}$ 
by a factor $\sim B/B_{\rm Q}$ is readily seen to be a consequence of
annihilation-creation equilibrium:  the cross section for the reverse
$e^\pm$ annihilation process is suppressed by a factor $\sim (B/B_{\rm Q})^{-1}$.  A simple case is
a thermal pair gas with a temperature $\lesssim m_ec^2$ and a 
density $\propto B/B_{\rm Q}$, this factor arising from the phase
space volume in the plane perpendicular to ${\bm B}$.

The magnetic field is weaker at greater distances from the magnetar, and we adopt the
non-magnetic cross section where $B < B_{\rm Q}$.
Selecting out a single polarization state (that is, the O-mode), one has \citep{Breit1934}:
    \ba
        \sigma_{\gamma\gamma}[B=0] = \pi r_e^2 \biggl[-SC^{-3} - \frac{3}{2} SC^{-5} - 
        \frac{3}{2} \theta C^{-6} \nn + 2\theta C^{-4} + 2\theta C^{-2}\biggr].
    \ea
Here, $\{S,C\} = \{\sinh{\theta},\cosh\theta\}$ and the COM photon energy is parameterized
as $[2/(1-\mu_{12})]^{1/2}\hbar(\omega_1\omega_2)^{1/2} = m_ec^2\cosh\theta$.  
Most of the pairs are created in the inner part of the magnetosphere where $B > B_{\rm Q}$;  
we do not represent the complicated transition between the two limiting expressions for $\sigma_{\gamma\gamma}$.

\vfil\eject
\subsection{Compensating Loss of Pairs}

    The production of pairs outside the gamma-ray emitting twist zones is compensated
    by (i) volumetric annihilation and (ii) loss through the surface of the star.  
    In addition, (iii) pairs are preferentially pushed away from the star on polar field lines,
    where the outward pressure of keV photons acting at the electron cyclotron resonance 
    provides a strong barrier to inward motion beyond a radius $(15-30)\,R_{\rm NS}$ 
    \citep{Thompson2008a, Beloborodov2013b}.
    Radiation pressure from the heated surface does not limit the impact and annihilation
    of positrons, at least when the radiation intensity is low enough that only a fraction of 
    gamma rays experience collisions. 

    The pair density $n_p$ we obtain is much larger than the corotation density $n_c = 
    |\bOmega\cdot{\bm B}|/2\pi ec$ when the rotation period is larger than a second,
    as typically observed in magnetars.  
    As a result, the plasma is locally charge balanced, $n_{e^+} \simeq n_{e^-}$.

\subsubsection{Volumetric Annihilation}

    In the idealized case of local kinetic equilibrium, there is a simple relation between the
    densities of pairs and gamma rays.   Annihilation into a single photon leads immediately to
    regeneration of the pair.  Some fraction of the time, the same is true for annihilation into two photons, 
    through the process of annihilation bremsstrahlung $e^+ + e^- \rightarrow \gamma + \gamma
    \rightarrow \gamma + e^+ + e^-$.  The cross section to annihilate into two photons, 
    both of which are below threshold for direct pair conversion ($\omega < \omega_{1\gamma}$) is \citep{Thompson2020}
        \ba\label{eq:sigann1}
        \sigma_{\rm ann,2\gamma}[B>B_{\rm Q}] = \frac{8 \pi r_e^2}{B/B_{\rm Q}} 
        \frac{\beta}{\gamma^2} \biggl[\frac{5}{9} - 1.29 \beta^2 - 0.0886 \beta^4 \nn 
        - \ln(\beta)\left( \frac{4}{3} + 0.936 \beta^2 + 4.54\beta^4 \right) \biggr].\nn
    \ea
Balancing the rate of pair creation by two photons of comparable energy
with annihilation of $e^\pm$ with kinetic energy $\hbar\omega-m_ec^2$ gives 
\be\label{eq:nratio}
{n_p\over n_\gamma} 
= \left({|1-\mu_{12}|\sigma_{\gamma\gamma}\over \sigma_{\rm ann,2\gamma}}\right)^{1/2} \simeq 
20\,\left({B\over 10\,B_{\rm Q}}\right).
\ee
(The coefficient in the right expression is for $\hbar\omega \sim 1.25\,m_ec^2$, 
corresponding to $\beta \sim 0.6$ for the created pair.)

The annihilation of $e^\pm$ is distributed along the confining magnetic field.  The net annihilation
rate within a flux rope of cross section $\delta A_\perp$ will be obtained in the approximation of pressure 
equilibrium along ${\bm B}$.  Given that $\delta A_\perp = (B/B_{\rm NS})^{-1} \delta A_{\perp,\rm NS}$
(here `NS' denotes the value where the rope connects to the star), the density profile is
    \be\label{eq:np_profile}
        n_{p} = n_{p,\rm NS} {\delta A_{\rm NS}\over \delta A} =
          n_{p,\rm NS}\frac{B}{B_{\rm NS}}.
    \ee
Consider first the part of a flux rope where $B > B_{\rm Q}$.    From Equation (\ref{eq:sigann1}),
the annihilation rate integrated across the cross section of the rope is distributed uniformly 
with respect to the longitudinal coordinate $\ell_\parallel = \int {\bm dx}\cdot\hat B$, 
\ba\label{eq:Ndot_ann}
{d(\delta\dot N_{\rm ann,2\gamma})\over d\ell_\parallel} &=& \delta A_\perp\,n_p^2\,\sigma_{\rm ann,2\gamma}(\beta,B)\,\beta c\nn
&=& \delta A_{\perp,\rm NS}\,n_{p,\rm NS}^2\,\sigma_{\rm ann,2\gamma}(\beta,B_{\rm NS})\,\beta c.\nn
\ea
This quantity is integrated along each flux tube from its intersection with the star out to a 
polar angle $\theta = \pi/2$.

Farther away from the surface of the star, where the magnetic field is weaker, the
annihilation cross section can instead be approximated by the vacuum expression
\citep{Berestetskii1982}:
    \ba\label{eq:sigann2}
        \sigma_{\rm ann,2\gamma}[B=0] &=& \frac{\pi r_e^2  (1-\beta^2)}{4 \beta} 
        \biggl[ \frac{3-\beta^4}{\beta} \ln\left(\frac{1+\beta}{1-\beta}\right) \nn
        && - 2(2-\beta^2) \biggr],
    \ea 
here defined in the COM frame.  We take the annihilation cross section $\sigma_{\rm ann,2\gamma}$ to 
be the minimum of expressions (\ref{eq:sigann1}) and (\ref{eq:sigann2}).

One sees that most annihilations are concentrated in the zone $B > B_{\rm Q}$ on flux ropes 
extending far enough that the magnetic field drops below $B_{\rm Q}$.

\subsubsection{Surface Annihilation}\label{sec:surface_ann}

Positrons also annihilate at the surface of the star. Above the surface, an atmosphere of ions and electrons
backscatters incident charged particles, reducing surface annihilation.
    The restriction of charged particle motion to one dimension (along ${\bm B}$) leads to 
    significant flux of reflected pairs:  only $1/3$ of those incident on the atmosphere are absorbed.
    Warm positrons and electrons both backscatter off ions with a cross 
    section $\sigma_{i\pm} = \pi r_e^2 / \beta^4\gamma^4$ \citep{Pavlov1976, Kostenko2019}, but 
    positrons backscatter more rapidly off atmospheric electrons through the 1-photon annihilation channel.
    The Coulomb cross section $\sigma_{i\pm}$ is smaller than  $\sigma_{\rm ann,1\gamma}/2$
    by a factor $\sim 0.06\,(B/10\,B_{\rm Q})$ (for a typical particle speed $\beta \sim 0.6$:
    see Equation (\ref{eq:sigma1gamma})).

    Annihilation takes place at greater depths than backscattering, but still at a moderate depth
    to O-mode electron scattering.  To see this,
    we note that annihilation into two photons (both below threshold for pair conversion)
    has a cross section $\sigma_{\rm ann,2\gamma} \ll \sigma_{\rm ann,1\gamma}$.  An absorbed positron 
    reaching an electron column $N_e$ below the surface has traversed a total column $N_e^2 \cdot\sigma_{\rm ann,1\gamma}/2$; it annihilates at a mean depth  
    \ba\label{eq:anndepth}
    \tau_O < \tau_{\rm T} &\sim& {\sigma_{\rm T} \over (\sigma_{\rm ann,2\gamma} \,
    \sigma_{\rm ann,1\gamma}/2)^{1/2}}\nn
    &=& 0.3\,\left({B\over 10\,B_{\rm Q}}\right).\qquad (\beta \sim 0.6)
    \ea

    The backscattered particle fraction is derived straightforwardly.  Given that scattering occurs on
    average at a column $N_s$ of target particles, the probability of first scattering at column $N_1$
    is $dP_1 = e^{-N_1/N_s}\,dN_1/N_s$, and the total probability to escape the atmosphere after
    this first scattering is $P_{1,\rm esc} = \int_0^{\infty} (dN_1/N_s) e^{-2N_1/N_s} = 1/2$.  
    The probability to escape after $2n$ more backscatterings is a factor $4^{-n}$ times smaller,
    leading to a total backscattering probability 
\be\label{eq:Pesc}
    P_{\rm esc} = \sum_{n=0}^\infty P_{2n+1,\rm esc} = {2\over 3}.
\ee
    Note that this result does not depend on the tilt of the magnetic field with respect to 
    gravity.

    The scattering of positrons back into the magnetosphere is primarly by warm electrons.  Note that
    collisions of warm and cold electrons do not change the one-dimensional electron
    distribution function, meaning that a population of warm electrons will be present in the
    outer atmosphere.  Warm electrons can backscatter off ions before cooling by bremsstrahlung
    emission, and so upward-moving warm
    electrons will have comparable density to downward moving warm electrons in the 
    shallow layer where positrons experience one-photon annihilation.

We must check that a positron does not first cool and annihilate in the atmosphere before backscattering.
    A positron can exchange energy and momentum with a cool atmospheric
    electron, given that the scattering is restricted to one spatial dimension.
    The resulting cool positron may either get reheated by a collision with a warm electron
    or instead annihilate (most likely with a cool atmospheric electron).
    Given a density $n_{e,c} \sim N_e/h$ of cool electrons in an atmosphere of scale height
    $h \sim k_{\rm B}T_{e,c}/m_pg$ and mean thermal speed $\beta_c
    \sim (k_{\rm B}T_{e,c}/m_ec^2)^{1/2}$,
    we may compare the rates of (i) annihilation into two photons (both below the threshold for
    single-photon pair creation) and (ii) backscattering off warm electrons with speed $\beta_h \sim 0.6$
    and  mean
    density $n_{e,h} = n_{p,\rm NS} \simeq \tau_{\rm T,\rm mag}/\sigma_{\rm T}R_{\rm NS}$.
     At a column $N_e \sim [\sigma_{\rm ann,1\gamma}(\beta_h)/2]^{-1}$
     typical of backscattering, we find
    \ba
    {\beta_c\,n_{e,c}\,\sigma_{\rm ann,2\gamma}(\beta_c)  \over 
    \beta_h\,n_{e,h}\,\sigma_{\rm ann,1\gamma}(\beta_h)/2}
    &\sim& {\alpha_{\rm em}^2\over \tau_{\rm T,\rm mag}}\left({B\over B_{\rm Q}}\right) 
    {m_pgR_{\rm NS}\over m_ec^2} \nn
    &=&0.013\,\tau_{\rm T,mag}^{-1}\left({B\over B_{\rm Q}}\right).
    \ea
    Note that $T_{e,c}$ and $\beta_c$ cancel from this expression.
    Atmospheric annihilation of positrons is therefore a second-order effect in our MC
    models, where $\tau_{\rm T,mag}\sim 0.3$ and $B = (4-10)B_{\rm Q}$. 

    The net result is that, counting only downward-moving positrons, we have a net surface annihilation rate
    at each end of a flux rope,
    \ba\label{eq:Ndot_s}
        \delta\dot N_s
        &=& \delta A_{\perp,\rm NS}\cdot {1\over 3}\cdot{1\over 2} n_{p,\rm NS} \beta c \nn
        &=& \frac{1}{9} \Rstar^2\,\delta\sqrt{3\cos^2\theta+1}\,\delta\phi \cdot n_{p,\rm NS} \beta c;
    \ea
    here, $\delta{X}$ represents the change in angular quantity $X$ across the flux bundle.

\subsubsection{Radiation Pressure Driven Outflow near the Poles}

    There is enhanced loss of plasma from the inner magnetosphere on field lines reaching
    far enough from the star for $e^\pm$ to resonantly scatter keV blackbody photons.
    Taking the photon and scattering particle to move radially near the magnetic poles, the resonant radius 
    corresponding to $\omega = \omega_{ce} = eB(r)/m_ec$ in the particle rest frame is
    \ba\label{eq:rres}
    {r_{\rm res}\over R_{\rm NS}} &=& \left[\gamma(1+\beta){eB_{\rm pole}\over m_ec \omega}\right]^{1/3}\nn
        &=&  30\,{B_{\rm pole,15}^{1/3}\over (\hbar\omega/{\rm keV})^{1/3}}\qquad (\beta = 0.6).
    \ea
    
    Dipolar field lines reaching this radius are anchored closer than an angle
    $\theta_{\rm res} \simeq (r_{\rm res}/R_{\rm NS})^{-1/2} \sim 10^\circ$ to the poles 
    (with surface magnetic flux density $B_{\rm pole}$).  Pairs moving along this polar 
    flux bundle feel a very strong drag when moving differentially with respect to the 
    photon field.  Because the pair kinetic energy flux below the resonance radius is much
    smaller than the X-ray Poynting flux, the pairs are forced rapidly outward in this zone.
    The corresponding particle loss is
  \be\label{eq:Ndot_res}
  \delta \dot N_{\rm res} = 
  \frac{1}{3} \Rstar^2\,\delta\sqrt{3\cos^2\theta+1}\,\delta\phi \cdot n_{p,\rm NS} 
       \beta c\quad (\theta < \theta_{\rm res}).
  \ee    

    The strength of this coupling may be gauged as follows.
    The radial force imparted to the pairs by a spherical flow of X-rays with luminosity $L_X$
    is $f_{\rm rad} = \gamma(1-\beta)(2\pi^2e^2/4\pi r_{\rm res}^2 m_ec^2)dL_X/d\omega$.
    Maintaining a radial pair velocity $\beta < 1$ within this photon field would require a large
    net energy input to the pairs:
    \be
    {f_{\rm rad}\cdot (1-\beta)r_{\rm res}\over m_ec^2} \sim  250\,
    {(dL_X/d\ln\omega)_{35}\over B_{\rm pole,15}^{1/3}(\hbar\omega/{\rm keV})^{2/3}},
    \ee
    evaluated here for $\beta \sim 0.6$.

\subsubsection{Balance between Pair Creation and Annihilation}

We may finally determine the equilibrium number of pairs on a given magnetic flux bundle from 
the solution to the following quadratic equation for $n_{p,\rm NS}$, obtained by summing the loss
terms from Equations (\ref{eq:Ndot_ann}), (\ref{eq:Ndot_s}) and (\ref{eq:Ndot_res}).
\be\label{eq:n_equil}
     \delta\dot N_p = 0 = \delta\dot N_{\gamma+\gamma\rightarrow e^++e^-} - 
     \delta\dot N_{\rm ann,2\gamma} - \delta\dot N_s - 
     \delta\dot N_{\rm res}.
\ee

\section{Magnetic Field and Currents}\label{sec:Bfield}

The changing X-ray pulse profiles of magnetars suggest the presence of
non-dipolar structure in the electric currents whose dissipation 
drives nonthermal X-ray emission (e.g. \citealt{Woods2001}).  The geometry of 
the underlying magnetic field is only weakly constrained.  
For our purposes, the two essential components are (i) a global
dipole magnetic field, part of which connects
to the outer magnetosphere and (ii) localized currents
connecting to crustal shear bands.  These localized currents are closely aligned with the field but generally 
do not flow near the magnetic poles;  they involve localized shearing of the field.  

Simple choices of the magnetic field and currents will suffice to demonstrate the effects 
of non-local $\gamma-\gamma$ pair creation.  The large-scale field may be weakly twisted but is assumed to
be predominantly dipolar.  The magnetic and particle density fields are represented in flat spacetime.  
Outside the star (radius $R_{\rm NS}$) and in spherical coordinates $(r,\theta,\phi)$,
   \be
   {\bm B}(r,\theta) = \bnabla \Psi\times\bnabla\phi  = {1\over r\sin\theta}\bnabla\Psi\times \hat\phi. \qquad (r \geq R_{\rm NS})
   \ee
   Poloidal magnetic flux surfaces are isocontours of the function
   \be 
   \Psi(r,\theta) = \sin^2\theta\, {{\mu}\over r} = {\mu\over{\cal R}}.
   \ee
   Here, $\mu$ is the magnetic moment of the star and ${\cal R} = r/\sin^2\theta = 
R_{\rm NS}/\sin^2\theta_\star$ is the largest radius reached by a dipolar field line
   that is anchored at polar angle $\theta_\star$.
   
\begin{figure}[ht]
        \includegraphics[width = \linewidth]{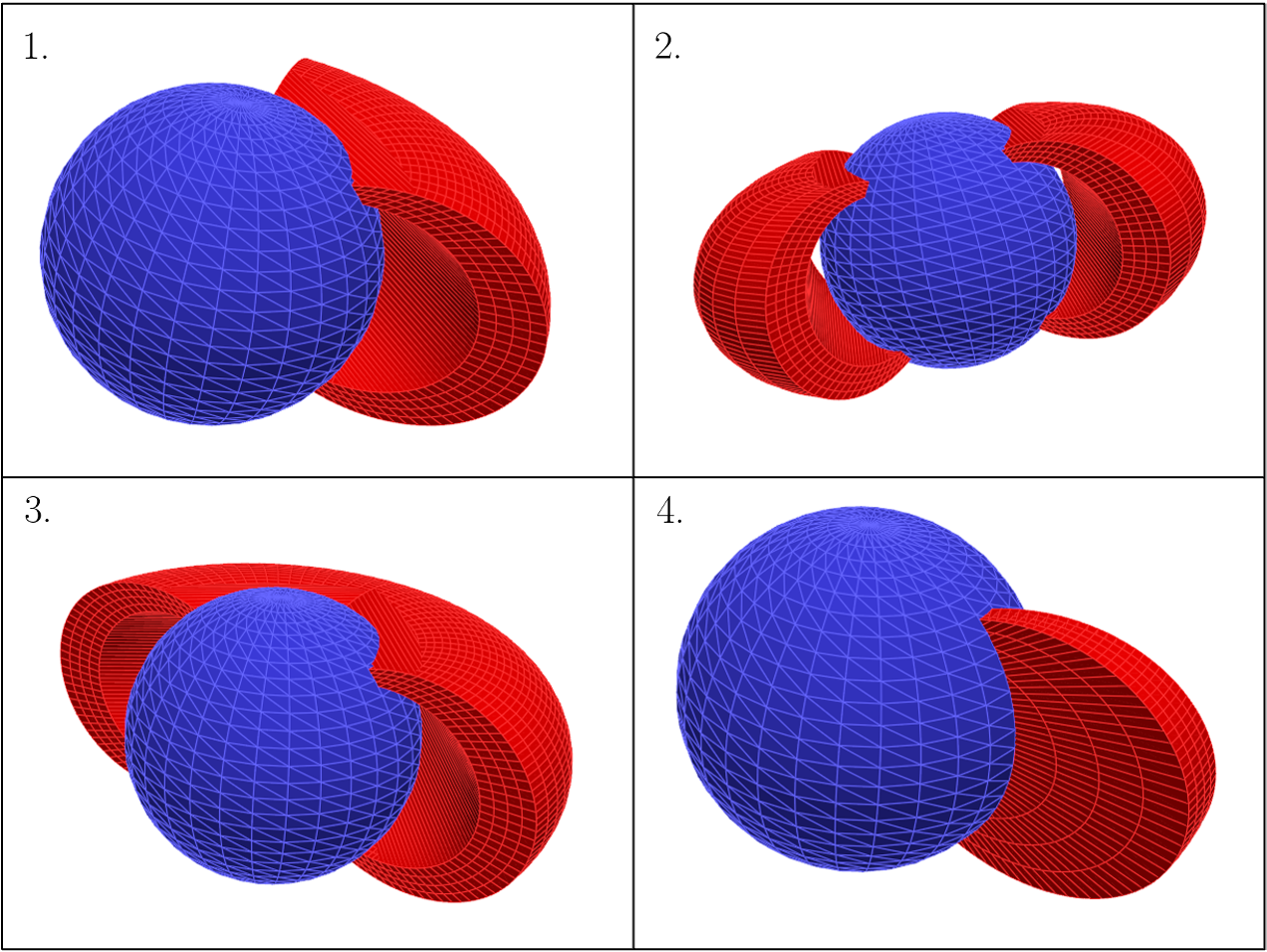}
        \caption{Hard X-rays and gamma rays are emitted from the surfaces of $e^\pm$-loaded magnetic 
        arcades (red) that are anchored to shear bands in the magnetar crust.  
        MC calculations of photon emission and collisions are carried out for the four cases shown. Top left:  
        Quarter shell extending over $\Delta\phi = \pi/2$.  Top right: Two antipodal quarter shells. Bottom left: 
        Half shell extending over $\Delta\phi = \pi$. Bottom right: thin wedge extending over $\Delta\phi = \pi/16$.  
        Parameters are listed in Table \ref{tab:structure_case}.}
        \label{fig:emission_structures}
    \end{figure}

We explore four different configurations in which the currents are confined to discrete arcade structures
(Table \ref{tab:structure_case} and Figure \ref{fig:emission_structures}).
   Electron-positron plasma, supporting the current and 
   emitting a hard spectrum of X-rays and gamma rays, is present in these twist zones.
   Because charged particles can flow along the magnetic field, but are confined by 
   radiative de-excitation to the lowest Landau state, these emission structures
   are defined by poloidal slices covering a discrete range of ${\cal R}$.  They also 
   cover a discrete range of azimuthal angle $\phi$.  In the case of the quarter shell
   structure (model 1), a mirror-symmetric arcade oppositely placed is also considered (model 2).
   Adding this second emission component widens the exposure to gamma rays 
   and induces stronger angular dispersion in overlapping rays.

 \begin{table}[ht]
        \begin{tabular}{l|lll}
             Case & $\theta_{\star, 1}$ & $\Delta \theta_\star$ & $\Delta \phi$\\ 
            \hline \hline 
             Quarter Shell & $\pi/4$ & 0.1234 & $\pi/2$ \\
             Half Shell & $\pi/4$ & 0.1234 & $\pi$ \\
             Thin Wedge & $\pi/4$ & $\pi/4$ & $\pi/16$ \\
        \end{tabular}
        \caption{Parameters of gamma ray emission structures. $\theta_{\star, 1}$ is the angle closest to the 
        pole where the emission surface intersects the star, $\Delta \theta_\star$ is the range of polar angle 
        covered by the emission structure, and $\Delta \phi$ is the range in azimuthal angle.  One case 
        (2 in Figure \ref{fig:emission_structures}) includes a second quarter shell structure that is offset by 
        angle $\pi$ in azimuth from the first.}
        \label{tab:structure_case}
    \end{table}
   
     \vfil\eject
       \subsection{Polar Magnetic Shear}

    Magnetars experiencing accelerated spindown appear to have inflated 
    magnetospheres\footnote{Related possibilities are that the polar field lines support 
    an enhanced persistent flux of Alfv\'en waves \citep{Thompson1998} or relativistic particles 
    \citep{Harding1999}.  Some fine tuning is required to launch the Alfv\'en wave or particle
    flow on a sufficiently narrow field bundle that it strongly amplifies the spindown, even while
    the hard X-ray continuum has already diminished \citep{Archibald2015}.}
    supporting a polar twist \citep{Thompson2002, Thompson2008b, Beloborodov2009}.
    One of our principal goals is to compare the density of pairs created non-locally 
    by $\gamma-\gamma$ collisions with the minimal (critical) density that sustains a polar twist.  
    When the global current can be supplied in this way, the polar voltage is greatly reduced.

    A fixed azimuthal offset $\Delta\phi_{\rm N-S}$ between the two magnetic hemispheres will generate a current
    \be
        \frac{4 \pi}{c} \mathbf{J} = \nabla \times \mathbf{B} 
        \simeq \Delta \phi_{\rm N-S} \frac{\mathbf{B}}{\Rmax}.
    \ee
    The critical space density of pairs $n_p = n_{e^+} = n_{e^-}$
    of charge $\pm e$ drifting with speed $\pm\beta c$ is $n_p = J/2|e|\beta c$.  Hence,
    \be \label{eq:twist_density}
        n_{\rm crit} = \frac{\Delta \phi_{\rm N-S}}{8 \pi \beta} \frac{B}{e\Rmax}.
    \ee 
    This density is usefully normalized by the Thomson cross section
    $\sigma_{\rm T}$ to give a characteristic optical depth that
    is not far off the optical depth to scattering for O-mode
    X-rays,
    \be
        2n_{\rm crit}\sigma_{\rm T}\Rstar = 
        \frac{2\Delta \phi_{\rm N-S}}{3 \beta} \alpha_{\rm em} 
        \left(\frac{B}{B_{\rm Q}}\right)\left(\frac{\Rstar}{\Rmax}\right).
        \label{eq:optical_depth}
    \ee 
Plasma of this minimal density is generally optically thin to electron-positron annihilation
and is too dilute to support the collisional state. 

     \begin{figure}[t]
        \includegraphics[width = \linewidth]{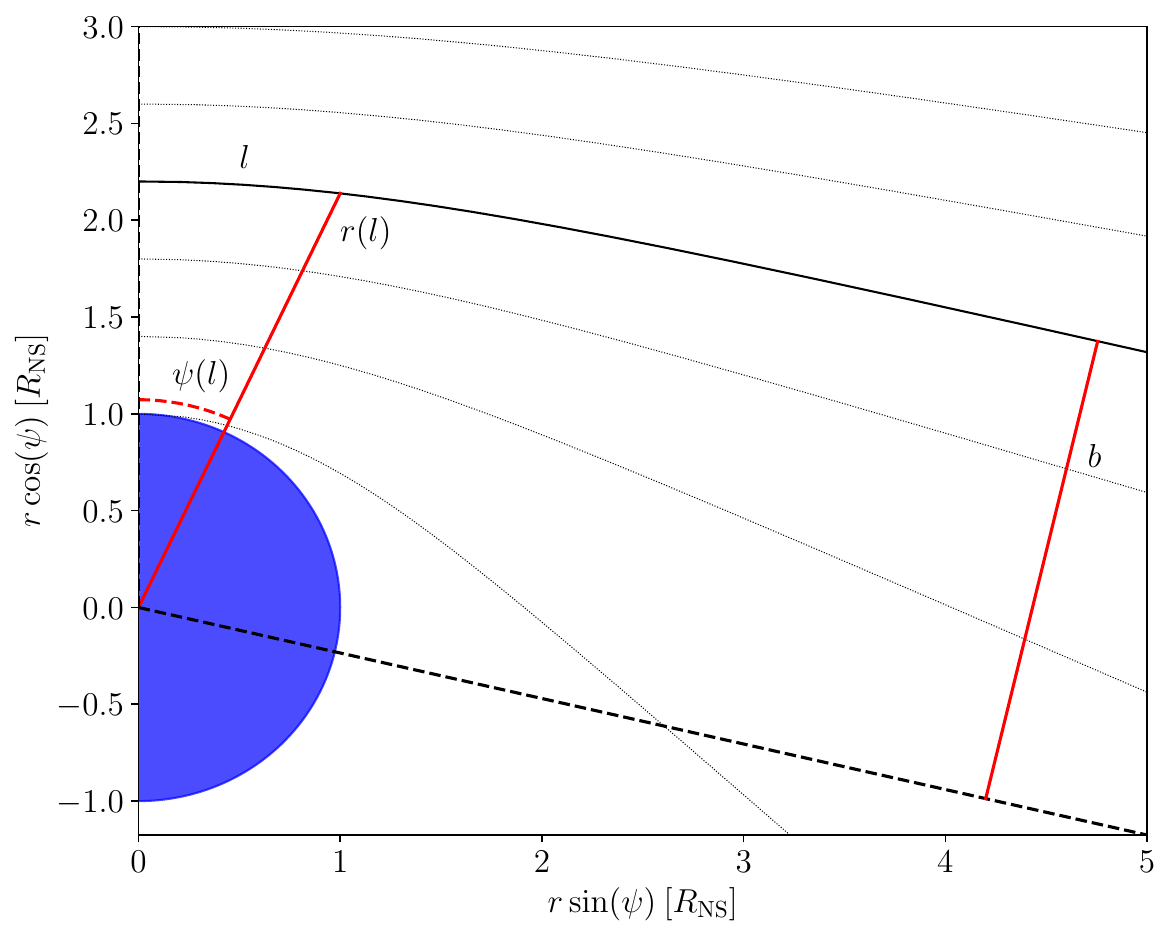}        
        \caption{Sample planar photon trajectories (black curves) around a non-rotating magnetar of 
        mass $1.4\,M_\odot$ and radius $\Rstar = 10$ km.   Coordinates $r$ and $\psi$ in the plane of the 
        geodesic are labelled by impact parameter $b$.  The geodesic library includes cases with 
        $\rmin < \Rstar$, which intersect the line $\Psi = 0$ with an angle smaller than $\pi/2$.}  
        \label{fig:traj_plane}
    \end{figure}

    \begin{figure}[ht]
        \includegraphics[width = \linewidth]{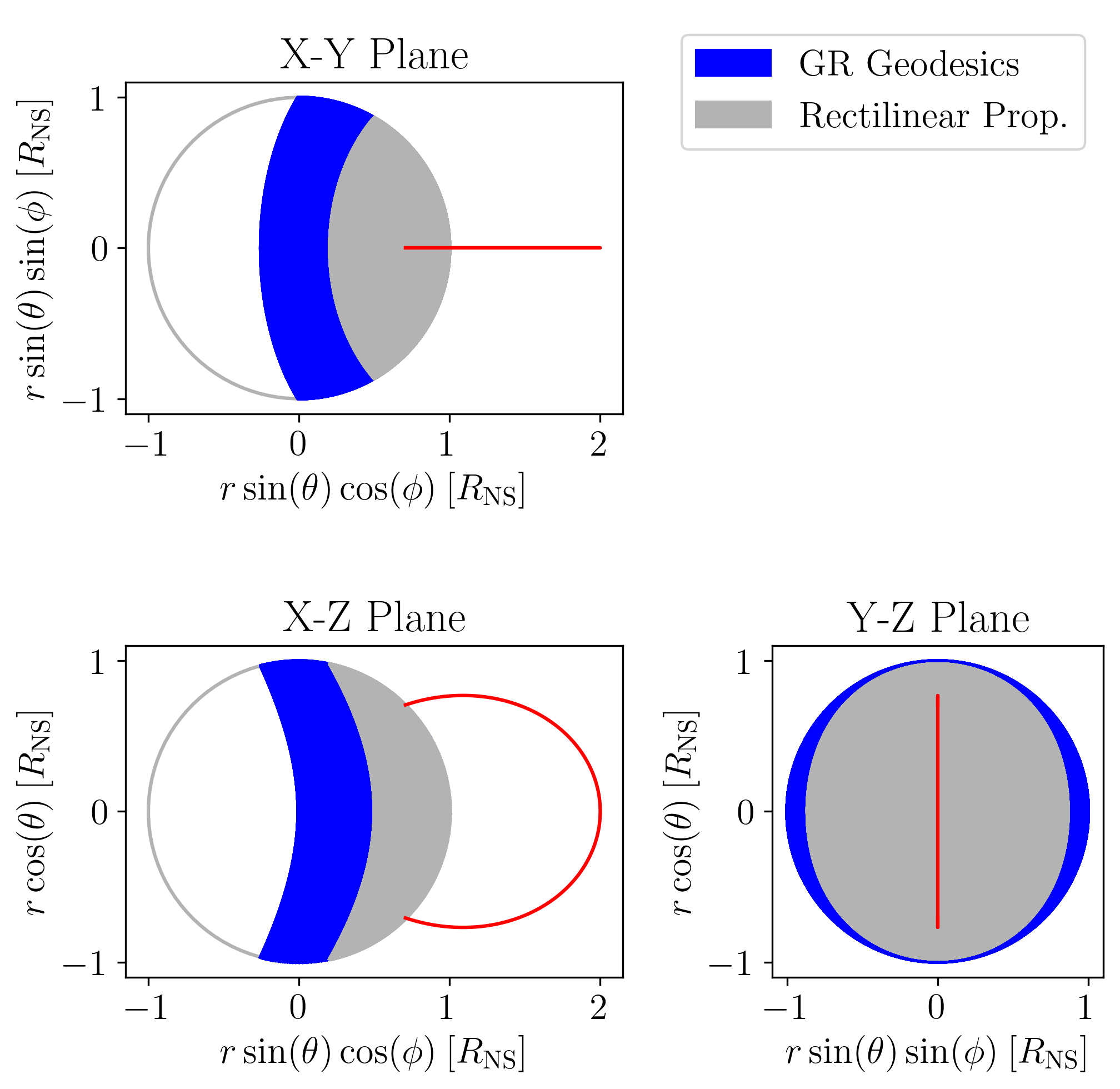}        
        \caption{Photons emitted from a slender fibril in a dipolar magnetic field
        can propagate through much -- but not all -- of the inner
        magnetosphere.  Some photons directly impact the surface.  The irradiated
        part of the surface is plotted in gray under the assumption of rectilinear
        propagation.  General relativistic geodesics are able to reach the expanded
        blue zone, thereby significantly increasing the irradiated portions of the
        stellar surface and surrounding magnetosphere.
        Here, the emitting fibril (red) is anchored at polar angle $\theta_\star = \pi/4.$}
        \label{fig:surface_heat}
    \end{figure}    

        \begin{figure}[ht]
        \vskip .4in
        \includegraphics[width = \linewidth]{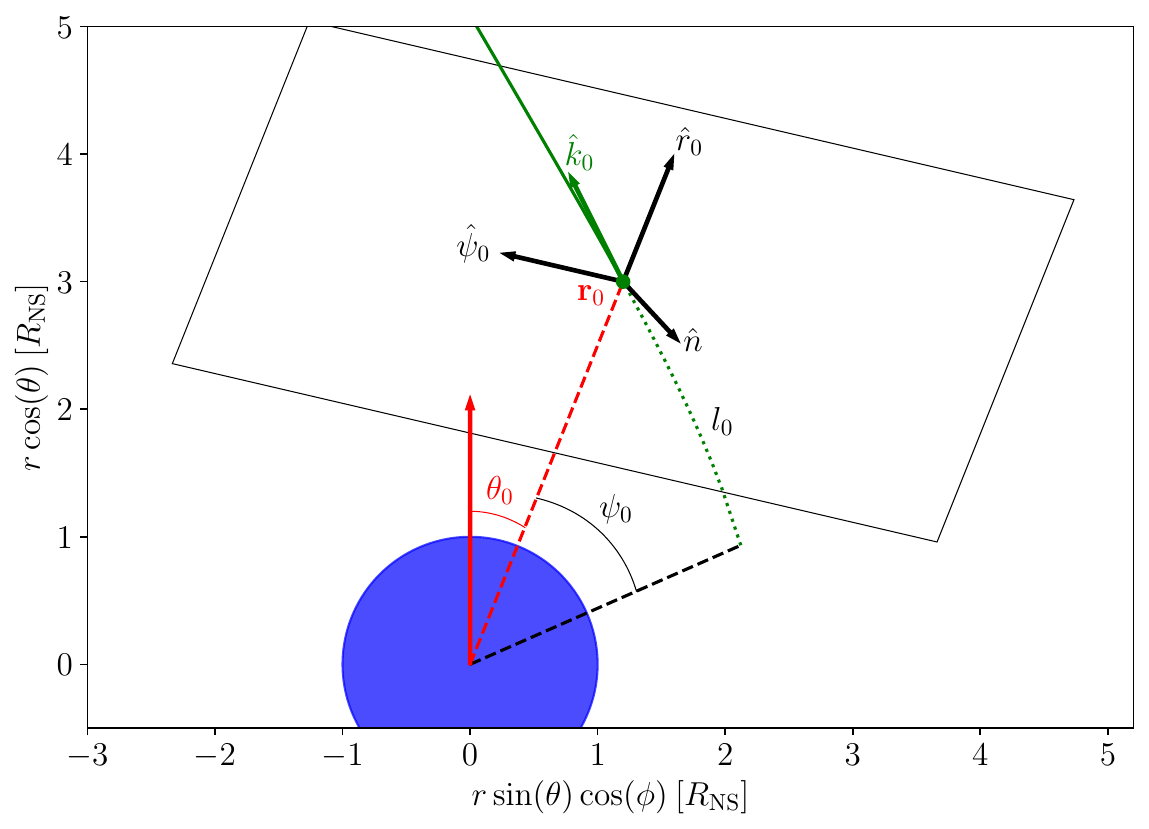}        
        \caption{Orientation of a sample geodesic plane in the frame of the star. The normal to the 
        plane is defined in terms of the emission point ${\bm r}_0 = (r_0, \theta_0, \phi_0)$ and 
        initial direction $\hat k_0$ by $\hat n = \hat k_0 \times {\bm r}_0 / |\hat k_0\times{\bm r}_0|$. 
         Variables in the geodesic plane are initial radius $r_0$
         and $\psi_0$, the angle between the emission point and the closest approach.}
        \label{fig:traj_sphere}
    \end{figure}

\section{Photon Propagation}\label{sec:trajectory}

  The curvature of photon trajectories is the principal general relativistic effect
  that must be taken into account to calculate the $e^\pm$ loading of the magnetic field
  outside the gamma ray emitting zones.  The refraction of O-mode trajectories due to a 
  varying index of refraction is subdominant to gravitational lensing when
  the surface field is of the strength considered here ($4-10\,B_{\rm Q}$).
  Some sample photon trajectories around a non-rotating star are shown in 
  Figure \ref{fig:traj_plane}.
  
  Lensing increases the volume that is exposed to pair-creating
  gamma rays if the emitting plasma is localized on one side of the star.
  To illustrate the importance of this effect, Figure \ref{fig:surface_heat} 
  shows the part of the stellar
  surface that is directly impacted by photons that are emitted in ally directions
  from the surface of a slender magnetic fibril.

    A photon trajectory outside a non-rotating star is a planar geodesic in Schwarzschild spacetime. 
    This curve can be parameterized by coordinates $(r, \psi)$ in a reference plane.  Our MC method,
    detailed in Section \ref{sec:mc}, builds a geodesic library that can be mapped onto general emission
    coordinates and directions.  Angle $\psi = 0$ is aligned with the closest 
    approach (radius $r_{\rm min}$) on rays that do not intersect the star.  Integrating from $\psi = 0$, a ray with impact parameter $b$
    is determined by (Figure \ref{fig:traj_plane}; \citealt{LLEM})
    \be
        \psi(r) = \int_{r_{\rm min}}^r \frac{d\psi}{dr} dr = \int_{r_{\rm min}}^r \frac{dr}{r^2} \left[ \frac{1}{b^2} - \frac{1}{r^2} \left(1 - \frac{r_g}{r} \right) \right]^{-1/2}.
    \ee
    Here, $r_g = 2GM/c^2$ is the gravitational radius.
    The minimum radius is related to the impact parameter by:
    \be 
        \rmin = 2\sqrt{\frac{b^2}{3}} \cos\left[\frac{1}{3}\arccos\left(\frac{-3 r_g}{2} \sqrt{\frac{3}{b^2}}\right)\right].
    \ee 

The geodesic plane is defined in the frame of the star by a choice of emission point 
${\bm r}_0$ and direction $\hat{k}_0$, as seen in Figure \ref{fig:traj_sphere}. The 
normal vector $\hat{n}$ and basis vectors are related by $\hat{r}_0$ and $\hat{\psi}_0$:
    \be\label{eq:basis}
        \hat{n} \equiv \frac{\hat{k}_0 \times \hat{r}_0}{|\hat{k}_0 \times \hat{r}_0|}; \quad \hat{\psi}_0 \equiv \hat{n} \times \hat{r}_0
    \ee 

Gravitational redshifting of the photon energy is readily included and is also
accounted for in our MC treatment.  From an emission point at radius $r_0$ to a pair
creation site (or surface impact) at radius $r_{\rm abs}$, a photon of initial energy $\omega_0$ redshifts or blueshifts to 
\be\label{eq:redshift}
\omega = \left[{g_{tt}(r_0)\over g_{tt}(r_{\rm abs})}\right]^{1/2}\omega_0
 \;=\; \left[{1-r_g/r_0 \over 1-r_g/r_{\rm abs}}\right]^{1/2}\omega_0.
 \ee
 The redshift is a $10\%$ correction, which can be significant because the
 broad emission spectrum cuts off sharply above energy $\sim m_ec^2$ when
 the effects of multiple scattering 
 and self-absorption in the emitting plasma are accounted for (see Equation \ref{eq:emspec} below
 and Figure 14 of \citealt{Thompson2020}).

Our MC treatment does not account for the general relativistic volume effect or the 
corresponding distortion of the background magnetic field.  Calculating the equilibrium pair
abundances depends on summing the pair creation rate over each dipolar magnetic flux 
element.  We choose a spatial grid aligned with the magnetic field, which is
easy for a flat-space dipole, but very complicated for more general field configurations.  

    \begin{table}[ht]
    \begin{tabular}{l|l}
    Parameter & Value \\
    \hline \hline 
    $\Rstar$    & 10 km \\
    $M_{\rm NS}$    &    1.4 $M_\odot$   \\
    $R_{\rm max}$   &  $32 \; \Rstar$    \\
    $N_{\Rmax}$ & 20 \\
    $N_\phi$ & 64 \\ 
    $N_\theta$ & 64 \\
    $N_{\theta k}$ & 8 \\
    $N_{\phi k, \rm max}$ & 8 \\ 
    $N_\omega$ & 128 \\
    $\hbar \omega_{\rm min}$ & 15 keV \\
    $\hbar \omega_{\rm max}$ & $2.5\,m_ec^2$ \\
    $\hbar \omega_0$ & $1.5\,m_e c^2$
    \end{tabular}
    \caption{Table of grid and emission parameters}
    \label{tab:parameters}
    \end{table}

\section{Monte Carlo Method}\label{sec:mc}

\begin{figure}[ht]
    \includegraphics[width = \linewidth]{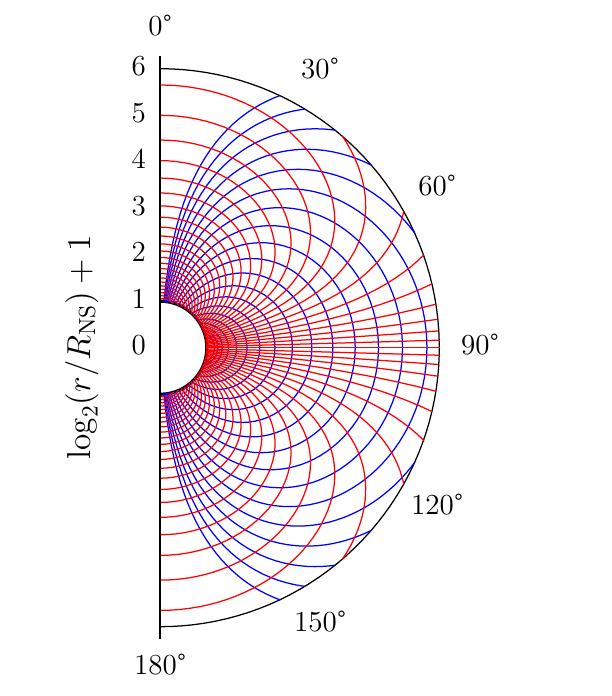}    
    \caption{Spatial grid aligned with flat-space dipolar magnetic field.  The field lines (blue) bounding
     spatial cells are curves of constant ${\cal R} = r/\sin^2\theta$.  The complementary set of boundary 
     lines (red) are curves of constant $\cos\varTheta = [1+\ln(r_{\rm pole}/r)]^{-1}$.}
    \label{fig:grid}
\end{figure}

    \subsection{Spatial and Angular Grids}\label{s:grid}
    
    The magnetospheric grid is aligned with the dipolar magnetic field (Figure \ref{fig:grid}), 
    in the expectation that the gamma-ray emitting surfaces are magnetic flux surfaces.
    Grid cells are bounded by surfaces of constant ${\cal R}$ (poloidal magnetic flux), 
    azimuthal angle $\phi$, and a conjugate angular variable chosen as 
    $\cos\varTheta = [1+\ln(r_{\rm pole}/r)]^{-1}$.  The isocontours of $\varTheta$ 
    are approximately orthogonal to those of ${\cal R}$; they are
    labelled by the radius $r_{\rm pole}$ at which they intersect the polar axis ($\theta = 0$),
    or equivalently by the angle $\widetilde\theta_\star$ at which they intersect the star. 
    The number of grid cells in each of these three coordinates is $(N_{\cal R}, N_{\varTheta}, N_\phi) 
    = (20, 64, 64)$, including both magnetic hemispheres.  
    
    The grid is uniformly spaced in $\phi$, in $\ln({\cal R}/\Rstar)$, and in $\widetilde\theta_\star^2$ 
    over the range $0 < \widetilde\theta_\star < \pi/2$.  It is symmetric on reflection between the north
    and south magnetic hemispheres, filling a spherical zone between radii $r = \Rstar$ and $r = R_{\rm max} = 32 \Rstar$.
    (Cells reaching $r = R_{\rm max}$ are cut off at that spherical radius.)
    Additional sets of cells cover the magnetic north and south polar axes; these are bounded by 
   ${\cal R} > {\cal R}_{g,\rm max} = 174\,R_{\rm NS}$ and by the variable $r_{\rm pole}$ (or equivalently
   $\widetilde\theta_\star$), but are not divided in $\phi$.

    The volume of a cell bounded by $(\Rmax_1, \Rmax_2)$, $(r_{\rm pole,1}, r_{\rm pole, 2})$, and 
    $(\phi_1, \phi_2)$ is given by
    \be
    V = (\phi_2 - \phi_1) \int_{\Rmax_1}^{\Rmax_2}  \Rmax^2 \int_{\theta_1(\Rmax)}^{\theta_2(\Rmax)} \sin^7 \theta d\theta d\Rmax 
    \ee
    where $\theta_{i}(\Rmax)$ is the polar angle at which the $\varTheta$ isocontour 
    labelled by $r_{{\rm pole}, i}$ intersects the flux surface labelled by ${\cal R}$.
    This is given implicitly by
    \be
    \frac{r_{{\rm pole}, i}}{\Rmax} = \left(1 - x_i^2 \right) e^{(1-x_i)/x_i}; \qquad
    \cos \theta_{i} \equiv x_i,
    \ee
    excepting when the intersection is implied to sit outside the computational domain
    (outer radius $R_{\rm max}$).
    
    The orientation of rays passing through each spatial cell is represented by an
   additional angular grid.  We choose spherical angular coordinates $(\theta_k, \phi_k)$, with 
   $\theta_k = 0$ aligned with $\theta = 0$ in each spatial cell.  The angular bins are uniformly divided in 
   $\theta_k$, with $N_{\theta k}=8$ rows.  The number of $\phi_k$-cells dividing each row of average
    polar angle $\bar\theta_k$ is given by $N_{\phi k} = \lceil N_{\phi k, \rm max} \sin(\overline{\theta}_k) \rceil$,
   with $N_{\phi k,\rm max}=8$.

    \subsection{Geodesic Library}

    Instead of solving the geodesic equation for each of billions of sample photons, we build 
     a reference library of planar Schwarzschild geodesics, 
     following the method described in
     Section \ref{sec:trajectory}.  These trajectories are labelled by impact
     parameter $b$ and uniformly distributed in $\log b$;  we also include the radial
     trajectory ($b=0$).   They are mapped directly into flat space
     and parameterized by the path length $l$. (Once again, our
     main goal in calculating geodesics is to account for blind
     spots in the photon density field produced by a localized
     emission structure;  the general relativistic corrections
     to path length and photon residency time are $O(r_g/r) \sim 10\%$.)
     
     Given a randomly chosen emission position ${\bm r}_0$ and
     ray tangent vector $\hat k_0$, we obtain the planar basis
     ($\hat n$, $\hat\psi_0$) from Equation (\ref{eq:basis}).
     From the emission radius $r_0$ and emission angle $\theta_{kr} = 
     \cos^{-1}(\hat k_0\cdot\hat r)$, we compute the impact parameter
     $b = r_0\sin\theta_{kr}(1-r_g/r_0)^{-1/2}$
     of the trajectory and then linearly interpolate between the two
     geodesics with the nearest $b$ values in the library.
     We then find the initial path length $l_0$, implicitly defined as $r(l_0) = r_0$, and angle $\psi_0 = \psi(l_0)$. 
     The trajectory $\mathbf{r}$ and ray tangent vector $\hat{k}$ 
     are easily computed as a function of the length $l$ of the
     embedded geodesic:
    \ba \label{eq:geodesic}
        \hat{r} &=& {{\bm r}\over r} = \cos[\psi(l) - \psi_0] \hat{r}_0 + \sin[\psi(l) - \psi_0] \hat{\psi}_0; \nn
        \hat{\psi} &=&  \hat n \times \hat r; \quad
        \hat{k} = \frac{dr}{dl} \hat{r} + r \frac{d\psi}{dl} \hat{\psi}.
    \ea 
    The endpoint of the trajectory corresponds to $r = R_{\rm max}$ or $r = R_{\rm NS}$, depending on
    whether it escapes the grid or first hits the star.

    \subsection{Photon Emission and Residency}

    The photon flux is assumed to be uniform across the surface of each emission structure and locally
    isotropic in direction.  The emission points of trial photons are therefore chosen randomly, with a
    uniform distribution by area. The initial $\hat k_0$ is defined relative 
    to the plane tangent to the emitting surface, being decomposed as
    $\hat k_0 = \cos\alpha\, \hat S + \sin\alpha\cos\beta\,\hat B + \sin\alpha\sin\beta\, \hat S \times \hat B$, where 
    $\hat S$ is the normal to the emitting surface. Isotropic emission corresponds to probability
    uniform in the direction cosine $\cos\alpha = \hat{k}_0 \cdot \hat{S}$ and uniform
    in $\beta$.   We only consider outward emission with $\cos\alpha > 0$.

    The source spectrum is chosen to represent annihilation bremsstrahlung emission from a collisional plasma,
    as modified by the effects of reabsorption and multiple electron scattering.
    Following the QED Monte Carlo results of \cite{Thompson2020}, we choose
    \be\label{eq:emspec}
    \frac{dL_{\gamma}}{d\omega} = c(n) \frac{L_0}{\omega_0} e^{-(\omega/\omega_0)^n},
    \ee 
    where $c(n)$ is a normalization constant, and $L_0$ is the total luminosity.
    The photon number spectrum $\hbar^{-1}dL_\gamma/d\omega$ is flat up close to 
    the energy $\hbar \omega_0$, above which it cuts off hyper-exponentially.  The parameters 
    $n=4$, $\hbar\omega_0 = 1.5\,m_ec^2$ fit the QED Monte Carlo spectra that correspond to a
    scattering depth $\sim 2-10$ across the source plasma.  The energies of trial photons 
    are drawn from the range
    $\{\hbar \omega_{\rm min} = 15\,{\rm keV}, \hbar \omega_{\rm max} = 3\,m_e c^2$\}, with
    $N_\omega = 32$ frequency bins spaced uniformly in $\log(\omega)$.  Photons that exceed
    the local pair conversion threshold energy $2m_ec^2/\sin\theta_{kB}$, either at emission or
    at some point on their trajectory, are assumed to be absorbed.

    Each photon trajectory intersects a certain number of grid cells. The change in path length across a given
    cell, $l_i \rightarrow l_f$, is determined by bisection.  A characteristic photon
    direction within the cell is $\hat{k} = [\hat{k}(l_f) + \hat{k}(l_i)]/2$. The photon also experiences gravitational 
    redshift according to Equation (\ref{eq:redshift}).
    The residency time in a cell is $t = (l_f-l_i)/c$;  this is summed separately 
    for each bin, as defined by $\omega$, $\hat k$ and cell number, to give a total residency time
    $\ttot$.

    After this procedure is completed for each photon (up to when it hits the star or escapes the grid),
    the photon number density distribution
    $n_{\gamma}(\hat k, \omega)$ is readily obtained.  From the total photon emission rate
    \be
    \Gamma_\gamma = \int \frac{1}{\hbar \omega} \frac{dL_{\gamma}}{d\omega} d\omega, 
    \ee 
    one has
    \be 
        n_{\gamma} \delta V = \frac{\ttot \Gamma_\gamma}{N_\gamma}. 
    \ee 
     Here, $\delta V$ is the cell volume and $N_\gamma$ the total number of trial photons.

     \subsection{Photon Collision Rate}

 The total pair production rate is determined in each grid cell
 by discretizing the integral (\ref{eq:ndotgg}) into a sum, 
     \ba
        \nprod &=& \delta V \sum_{\hat k_1, \hat k_2} \sum_{\omega_1, \omega_2}
        n_1(\hat k_1, \omega_1) n_2(\hat k_2, \omega_2)\nn
        && \times |1- \mu_{12}| c \, \sigma(\omega_1, \omega_2, \hat k_1, \hat k_2, B).
        \label{eq:n_product}
    \ea
    We then sum the pair production along each flux bundle.
    In the approximation that plasma drift is fast enough to equilibrate the pressure, according to Equation
     (\ref{eq:np_profile}), the creation of pairs is balanced against loss due to outflow, surface collision, 
    and annihilation using Equation (\ref{eq:n_equil}).  This gives the equilibrium surface density 
    $n_{p,\rm NS}$ and the magnetospheric density following Equation (\ref{eq:np_profile}).

    To ensure that photon collisions are not overcounted, we sort and bin
    the summed pair creation rate by the energy carried by the more energetic photon.
    Gamma rays with energies exceeding $2m_ec^2$
    see a relatively large absorption optical depth, as pair conversion 
    with photons of energy $\ll m_ec^2$ is kinematically allowed
    (Equation (\ref{eq:threshold})) and has a higher cross section (Equation
    (\ref{eq:sigmagg})).  We apply a conservative cut on $\nprod$ by 
    removing collisions in photon energy bins where the implied absorption rate
    (of the more energetic photon) is larger than its production rate.
    Examples are shown in
    the Appendix.  The net effect is that pair creation is dominated by
    collisions in which the higher photon energy lies in the range $(1-2)m_ec^2$.
    Photons of this energy also experience only limited electron scattering (Section \ref{s:npair}):  they are
    outside the resonance that enhances electron scattering at photon energies just below threshold
    for single-photon pair conversion ($\omega \lesssim \omega_{1\gamma}$; \citealt{Kostenko2018}).

     This bound on pair creation rate is slightly relaxed for arcade model 4 (the thin slice geometry).
      Here we find that about one half of the pair creation is concentrated in the cells adjacent
      to the emitting surface and, furthermore, that this photon absorption is mainly balanced by re-emission
      through volumetric $e^\pm$ annihilation.  The net effect is to displace outward slightly the
      emission surface without changing its overall geometry.  In model 4, we therefore allow the
      photon absorption rate to be up to 1.5 times the emission rate;  in practice, this only affects
      the highest energy bin for which photon collisions are counted.

   \begin{figure*}[ht]
    \centering
    \includegraphics[width = 0.7\linewidth]{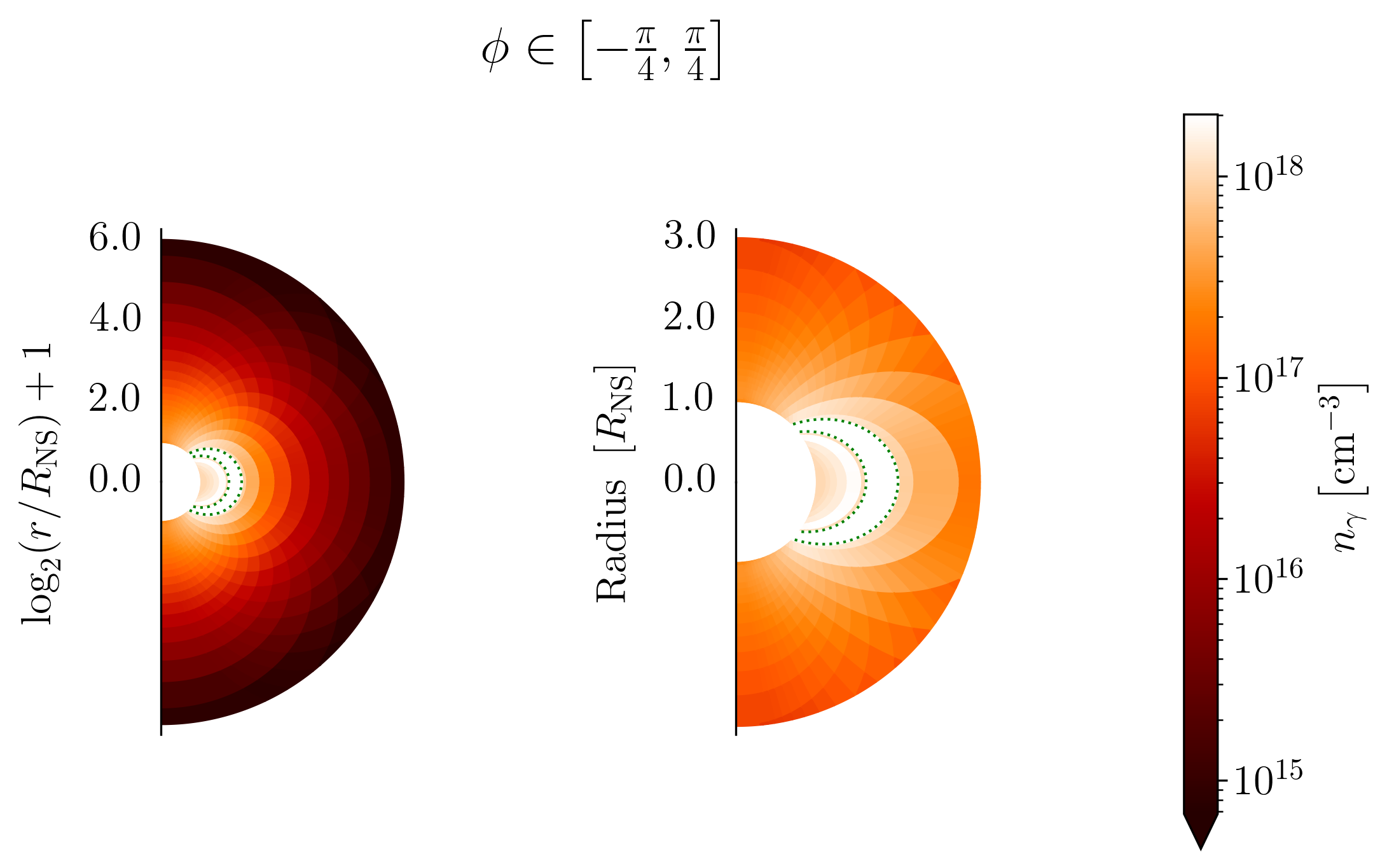}
    \caption{Photon density $n_\gamma$ obtained by MC sampling of uniform and isotropic emission from the
    surface of the quarter shell arcade (depicted in white, extending to
    twice the radius of the star).  Gamma ray luminosity 
    $L_0 = 10^{35}$ erg s$^{-1}$, with spectral parameters 
    $\hbar\omega_0 = 1.5\,m_ec^2$ and $n=4$ in Equation (\ref{eq:emspec}).  Plotted value of $n_\gamma$
    is averaged over 
    $\phi$ in the same quadrant from which gamma rays are emitted.  Left panel: full MC domain 
    (log radius);  right panel: zoomed in to $r < 3\Rstar$ (linear radius).}
    \label{fig:photon_density}
   \end{figure*}
   \begin{table*}
        \begin{ruledtabular}
        \begin{tabular}{c|cccccc}
             Model & $\gamma$ Conv.  & Surf. Ann. & Vol. Ann.  & Surf. Lum. [erg s$^{-1}$] & Max Temp [K] & $A_e / A_{\rm NS}$\\ 
             ($B_{\rm pole} = 10\,B_{\rm Q}$; $L_{0,35}=1$) & & & & & & \\
             (a) & (b) & (c) & (d) & (e) & (f) & (g)\\
            \hline \hline 
             1. Quarter Shell  &  12\,\% &  9.9\,\% &  2.9\,\%  & $1.1 \times 10^{34}$ & $2.8 \times 10^6$ &  7.1\,\%\\
             2. Double Quarter  & 8.9\,\% &  8.4\,\%&  1.0\,\%   & $9.7 \times 10^{33}$ & $2.3 \times 10^6$ &  15\,\%\\
             3. Half Shell & 11\,\%& 9.6\,\%&  1.9\,\%   & $1.0 \times 10^{34}$ & $2.5\times 10^6$ &  12\,\%\\
             4. Thin Wedge &  19\,\%& 12\,\%&  8.2\,\%  & $1.2 \times 10^{34}$ &$3.7 \times 10^6$ &  2.6\,\%\\
             5. Thin Wedge* &  11\,\% &  8.3\,\% &  3.0\,\%  & $8.0 \times 10^{33}$ & $3.1 \times 10^6$ &  3.7\,\%\\
            \hline\hline
             Model & $\gamma$ Conv.  & Surf. Ann. & Vol. Ann. & Surf. Lum. [erg s$^{-1}$] & Max Temp [K]&$A_e / A_{\rm NS}$\\ 
             ($B_{\rm pole} = 4\,B_{\rm Q}$; $L_{0,35}=1$) & & & & & & \\
             (a) & (b) & (c) & (d) & (e) & (f) & (g)\\
            \hline \hline 
             1. Quarter Shell &  5.2\,\% &  4.7\,\% &  0.7\,\%  & $7.5 \times 10^{33}$ & $2.5 \times 10^6$ &  7.7\,\%\\
             2. Double Quarter &  3.9\,\%&  3.8\,\%& 0.2\,\%   & $7.0 \times 10^{33}$ & $2.1 \times 10^6$& 16\,\%\\
             3. Half Shell & 4.8\,\%& 4.5\,\%& 0.4\,\%   & $7.4 \times 10^{33}$ & $2.2 \times 10^6$& 13\,\%\\
             4. Thin Wedge&  9.2\,\%& 6.8\,\%& 2.8\,\%   & $8.9 \times 10^{33}$ &$3.3 \times 10^6$&  3.0\,\%\\   
             5. Thin Wedge*& 5.4\,\%&  4.7\,\%& 1.0\,\%   & $5.9 \times 10^{33}$ &$2.8 \times 10^6$&  3.9\,\%\\
        \end{tabular}
        \end{ruledtabular}
        \caption{Quantities listed, for each emission model (a):  efficiency of conversion
        of gamma ray energy to pairs (b), fraction of energy lost to annihilation at the surface
        (c), fraction returned to gamma rays through volumetric annihilation (d), total 
        luminosity absorbed at the surface through annihilation and photon impact (e), peak surface effective 
        temperature (f), and the area $A_e$ containing half the absorbed surface luminosity (g).  
        Top and bottom tables show results for (i) polar magnetic field $10\,B_{\rm q}$ and 
        luminosity $L_0 = 1 \times 10^{35}$ erg s$^{-1}$ and (ii) magnetic field
        $4\,B_{\rm Q}$ and luminosity $L_0 = 1\times 10^{35}$ erg s$^{-1}$.
        Annihilation energy in (c) includes change in gravitational energy ($\beta \rightarrow 0.6$ at surface).
        Thin Wedge$^*$ show results after excising one layer of grid cells adjacent to the emission surface.}
        \label{tab:Temperature}
    \end{table*}

    \begin{figure*}
        \centering
        \includegraphics[width = 0.48\linewidth]{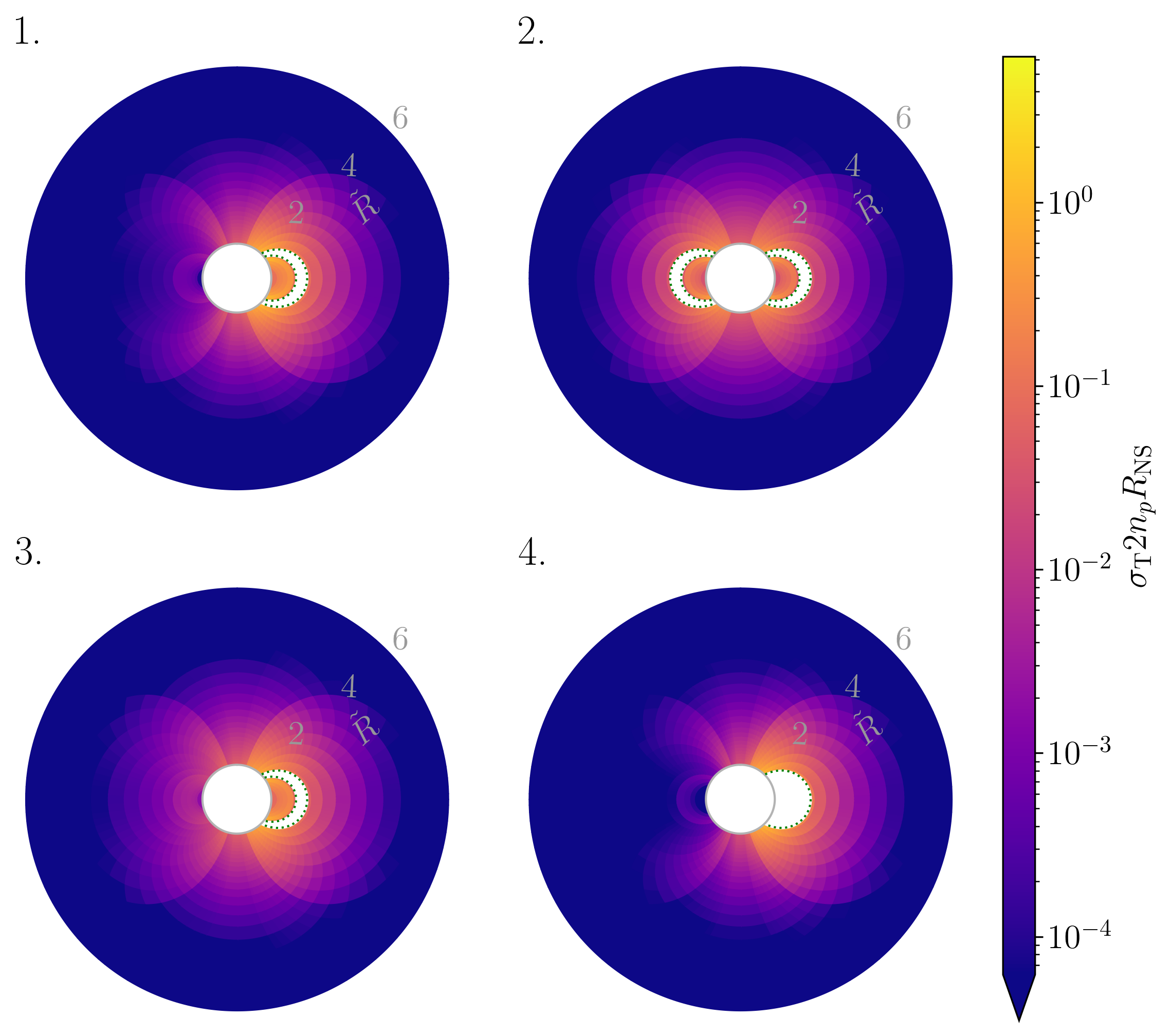} \hskip .2in
        \includegraphics[width = 0.48\linewidth]{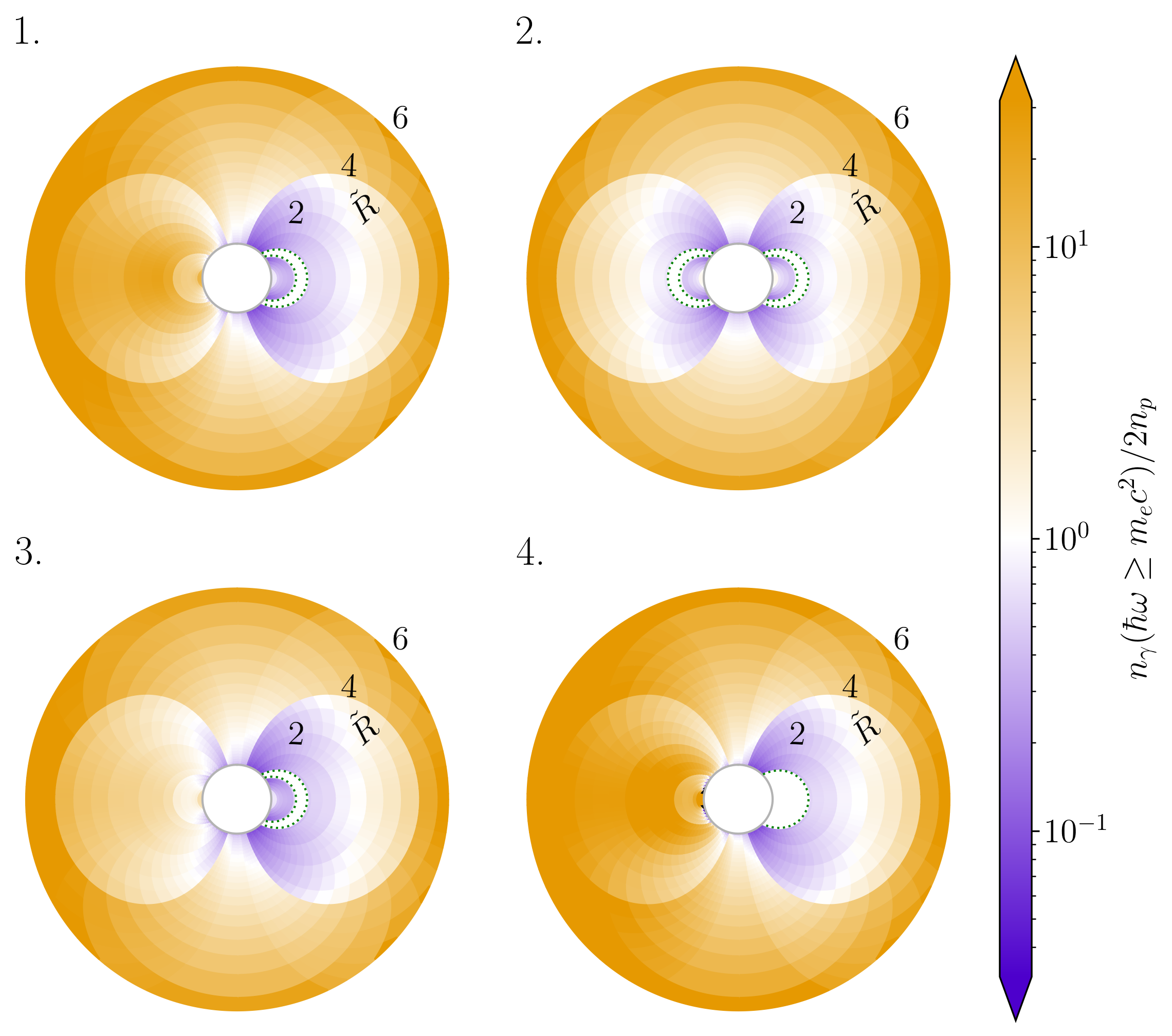}
        \caption{Left panel: Poloidal slice of the equilibrium density  $n_{e^+} + n_{e^-}$, normalized by 
        $(\sigma_{\rm T} \Rstar)^{-1}$, for all four types of emission structure (marked in white)
        and for $B_{\rm pole}=10\,B_{\rm Q}$, $L_0= 10^{35}$ erg s$^{-1}$.
        Right panel:  ratio of density of pair-creating photons to pair density.
        Values plotted are an average over $\phi$ in the same quadrant from which 
        gamma rays are emitted. Pair creation via $\gamma + \gamma \rightarrow e^+ + e^-$ is integrated over the 
        entire volume of a magnetic flux tube and balanced with volumetric annihilation, loss through the surface, 
        and loss across the keV electron cyclotron resonance surface
        (on field lines extending beyond $20(B/10\,B_{\rm Q})^{1/3}$ stellar radii).  
        Radial density distribution along a flux tube is given by Equation (\ref{eq:np_profile}).
        The radial variable is $\widetilde R \equiv \log_2{(r/\Rstar)} + 1$.}
        \label{fig:equil_slice}
    \end{figure*}        

    \begin{figure*}
       \centering
        \includegraphics[width = 0.5\linewidth]{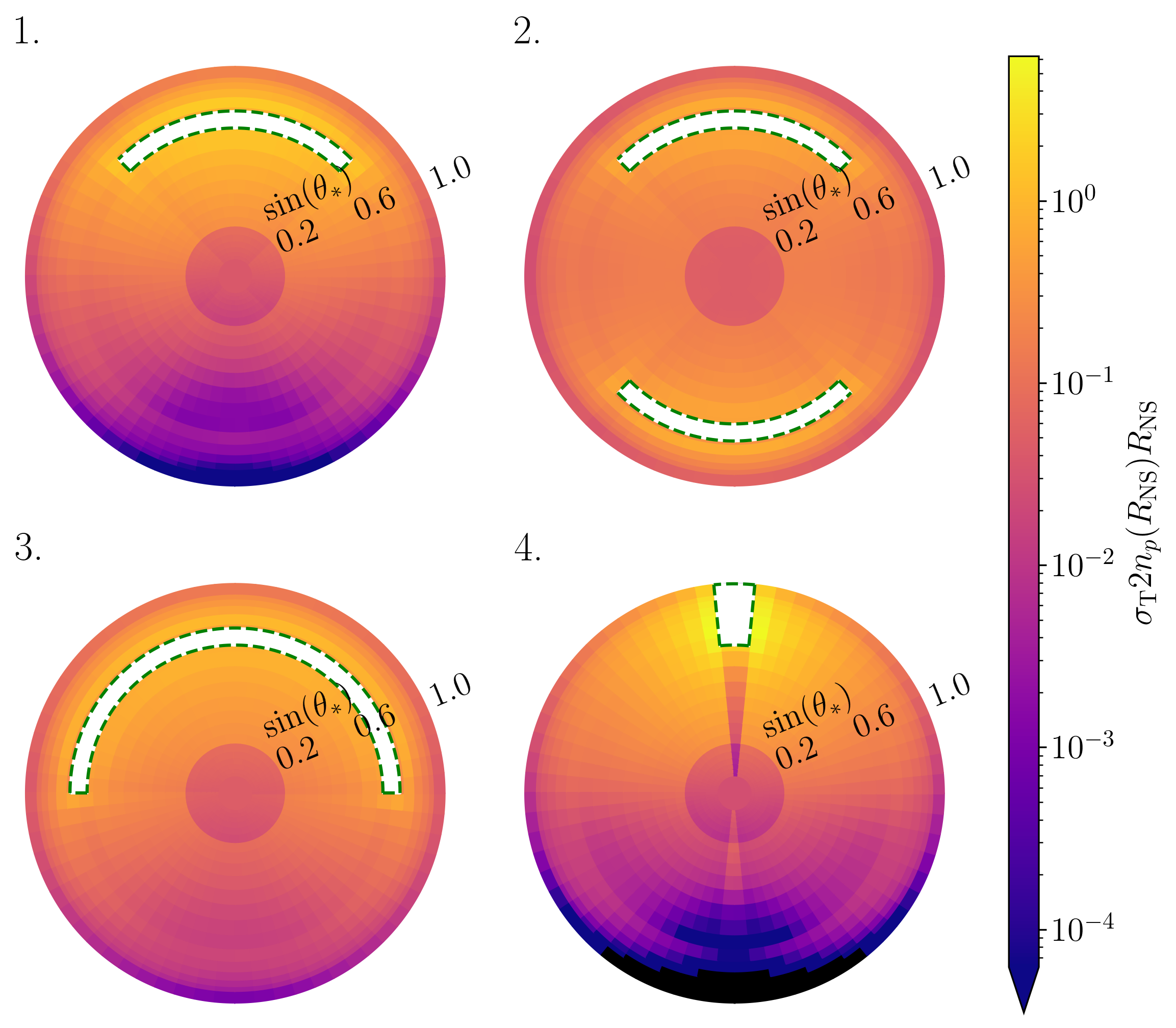}
        \caption{Equilibrium density $n_{e^+} + n_{e^-}$ at the surface of the star, normalized by 
        $(\sigma_{\rm T} \Rstar)^{-1}$, for all four types of emission structure 
        (marked in white) and for $B_{\rm pole} = 10\,B_{\rm Q}$, $L_0 = 10^{35}$ erg s$^{-1}$.
        View is along along a polar
        magnetic axis, with radial coordinate $\sin\theta_\star$ and azimuthal coordinate $\phi$.}
        \label{fig:equil_dens}
    \end{figure*}

    \begin{figure*}[t]
        \centering
       \includegraphics[width = 0.9\linewidth]{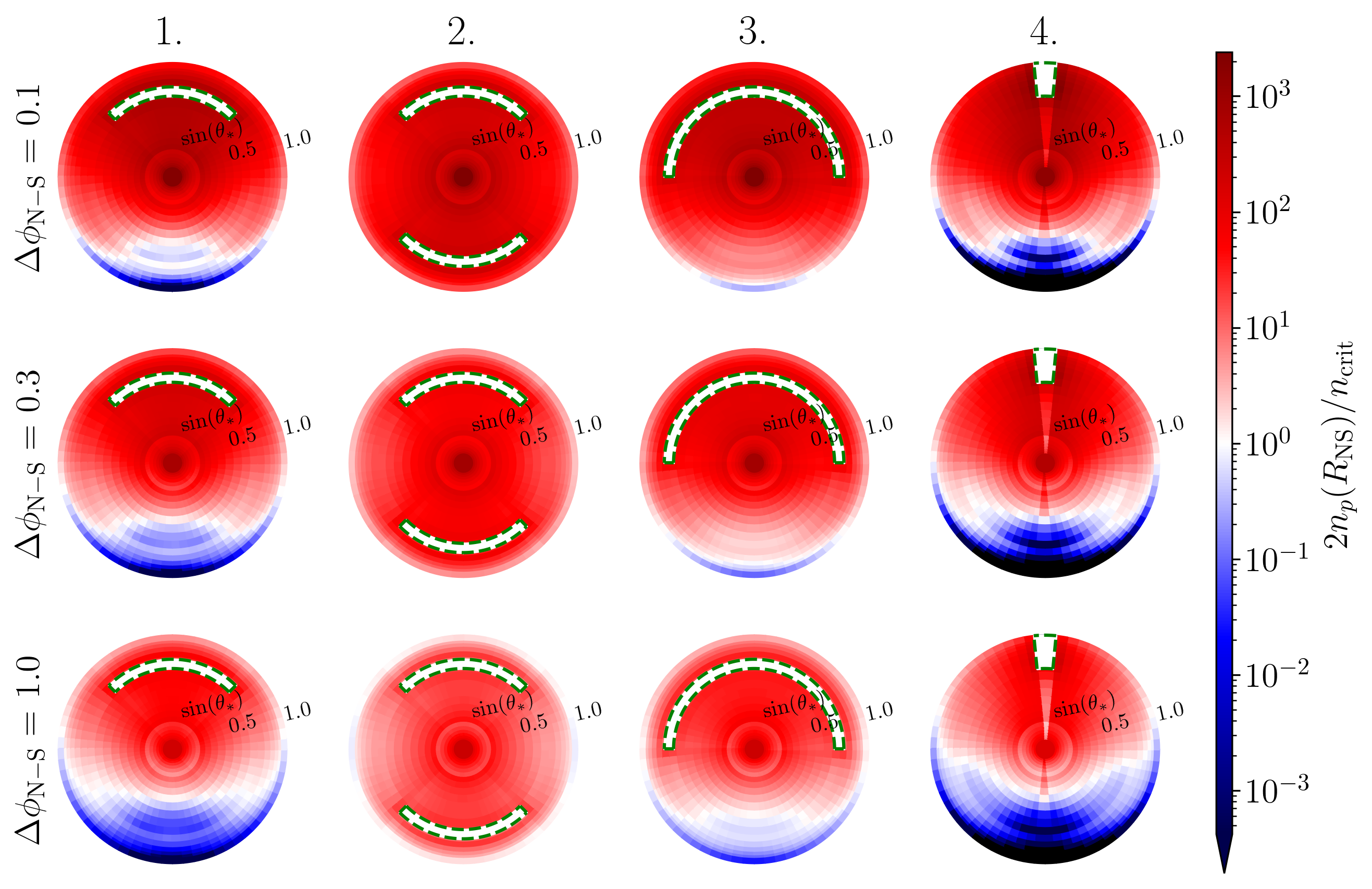}
        \caption{Comparison of the plasma density created by non-local pair creation,
         $\gamma + \gamma \rightarrow e^+ + e^-$, to the critical plasma density that is needed to support a global
        magnetospheric twist of angle $\Delta\phi_{\rm N-S}$ (Equation \ref{eq:twist_density}).  The result is
        shown for a range of twist angles;  a twist angle $\Delta\phi_{\rm N-S} = 1$ corresponds to 
        a substantial increase in spindown rate, by a factor $\sim 10$ \citep{Thompson2002}.}
      \label{fig:twist_ratio}
    \end{figure*}

\section{Results}\label{sec:results}

   Localized currents dissipating at a higher rate than their
   surroundings can be a copious source of low-energy gamma rays near a magnetar.   We present two sets of 
   MC runs, both with luminosity $L_0 = 1\times 10^{35}$ erg s$^{-1}$, 
   and with (i) polar magnetic field $B_{\rm pole} = 10\,B_{\rm Q}$ and  
   (ii) $B_{\rm pole} = 4\,B_{\rm Q}$,
      each involving $5.2\times 10^8$ trial photons in each of the four sample emission structures.
   Figure \ref{fig:photon_density} shows the 
   photon density field surrounding a quarter shell arcade (model 1 in Figure \ref{fig:emission_structures}).
   
   This radiative output, which is dominated by gamma rays of energy $\sim 0.5-1$ MeV, is modest by the 
   standards of quiescent magnetars.  When the hard X-ray photon index is $\Gamma_X = -1$ (as  
   adopted here), the output in the $15-60$ keV band is only $10^{34}\,L_{0,35}$ erg s$^{-1}$;
   this drops to $\sim 0.5\times 10^{34}\,L_{0,35}$ erg s$^{-1}$ when $\Gamma_X$ rises to $-0.6$,
   as measured in 1E 2259$+$586 and as inferred for pair annihilation in magnetic fields around $B_{\rm Q}$
   \citep{Thompson2020}.  

  The main results can be summarized as follows.
  \vskip .1in
  \noindent
  1. The pairs in equilibrium are more numerous than gamma rays in the
  inner magnetosphere, as seen in Figure \ref{fig:equil_slice}.  In part, this is because photon collisions have 
  a higher cross section than pair annihilation, $\sigma_{\gamma\gamma}/\sigma_{\rm ann,2\gamma} \sim
  (B/B_{\rm Q})^2$, and in part because there is significant reflection of downward moving pairs 
  from the magnetar surface (Equation \ref{eq:Pesc}).  
  \vskip .05in
  \noindent
  2. The pairs created above the magnetic poles can easily supply the current 
  associated with a weaker global magnetic twist, significantly limiting the polar voltage.
  \vskip .05in  
  \noindent 
  3.  A moderate fraction of the emitted gamma ray power is directly and indirectly absorbed 
  by the magnetar, creating an inhomogeneous surface radiation profile.  The large-scale magnetic
  field indirectly channels gamma ray energy toward the magnetar in the form of a cloud of trans-relativistic
  pairs, which dominate surface heating in comparison with the direct impact of gamma rays.
  \vskip .05in
  \noindent
  4.  A large optical depth to electron cyclotron scattering is produced around the
  resonance radius $r_{\rm res} \sim 20\,(B/10\,B_{\rm Q})^{1/3}\, R_{\rm NS}$
  by the outward pressure of keV photons on the pairs flowing near the magnetic poles.
  Rescattering in this zone will have a major impact on X-ray pulse profiles and X-ray 
  polarization, but less on the X-ray spectrum given the small net kinetic energy of the pairs.

    \begin{figure*}
        \centering
        \includegraphics[width = 0.48\linewidth]{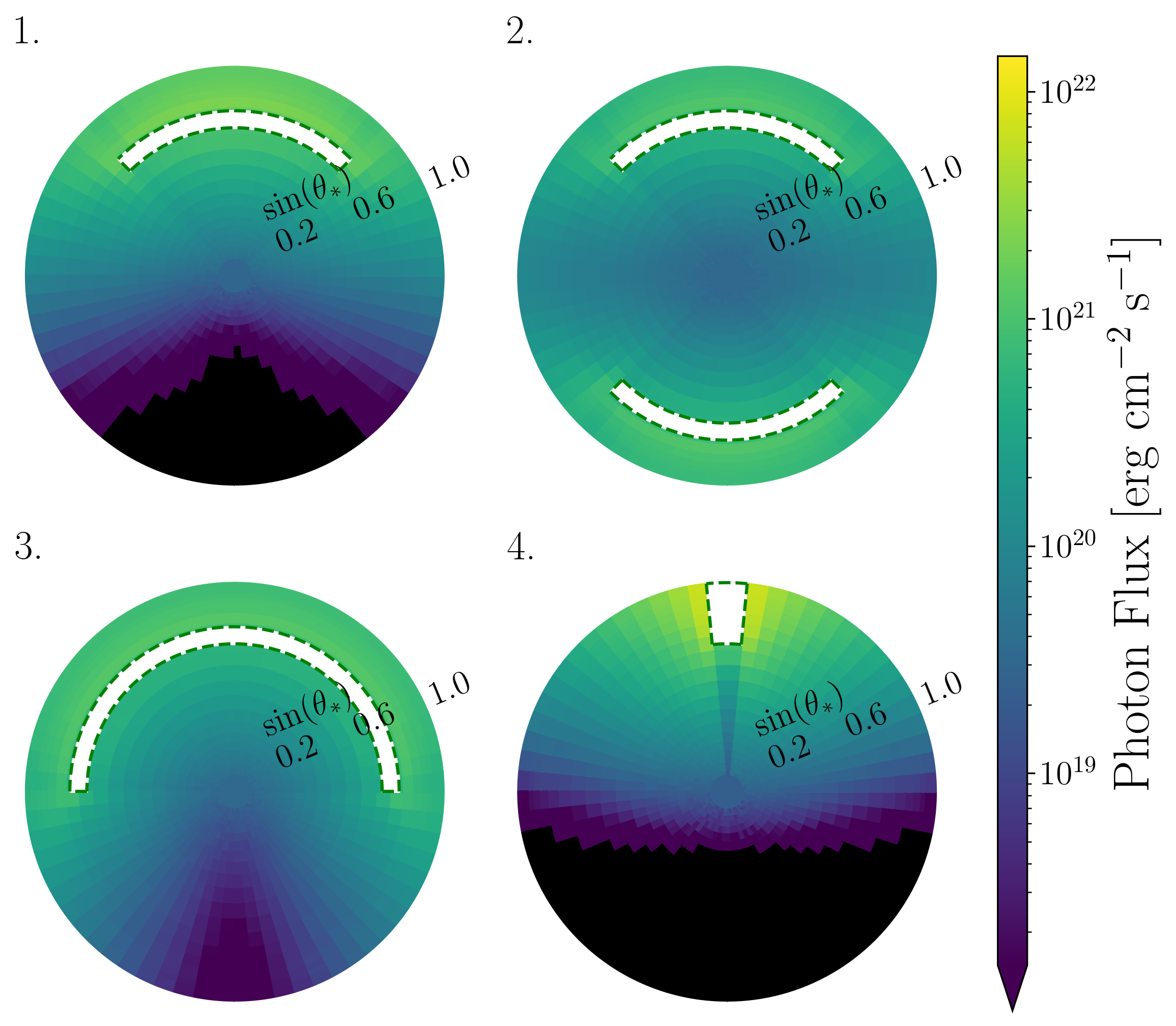}\hskip .2in
        \includegraphics[width = 0.48\linewidth] {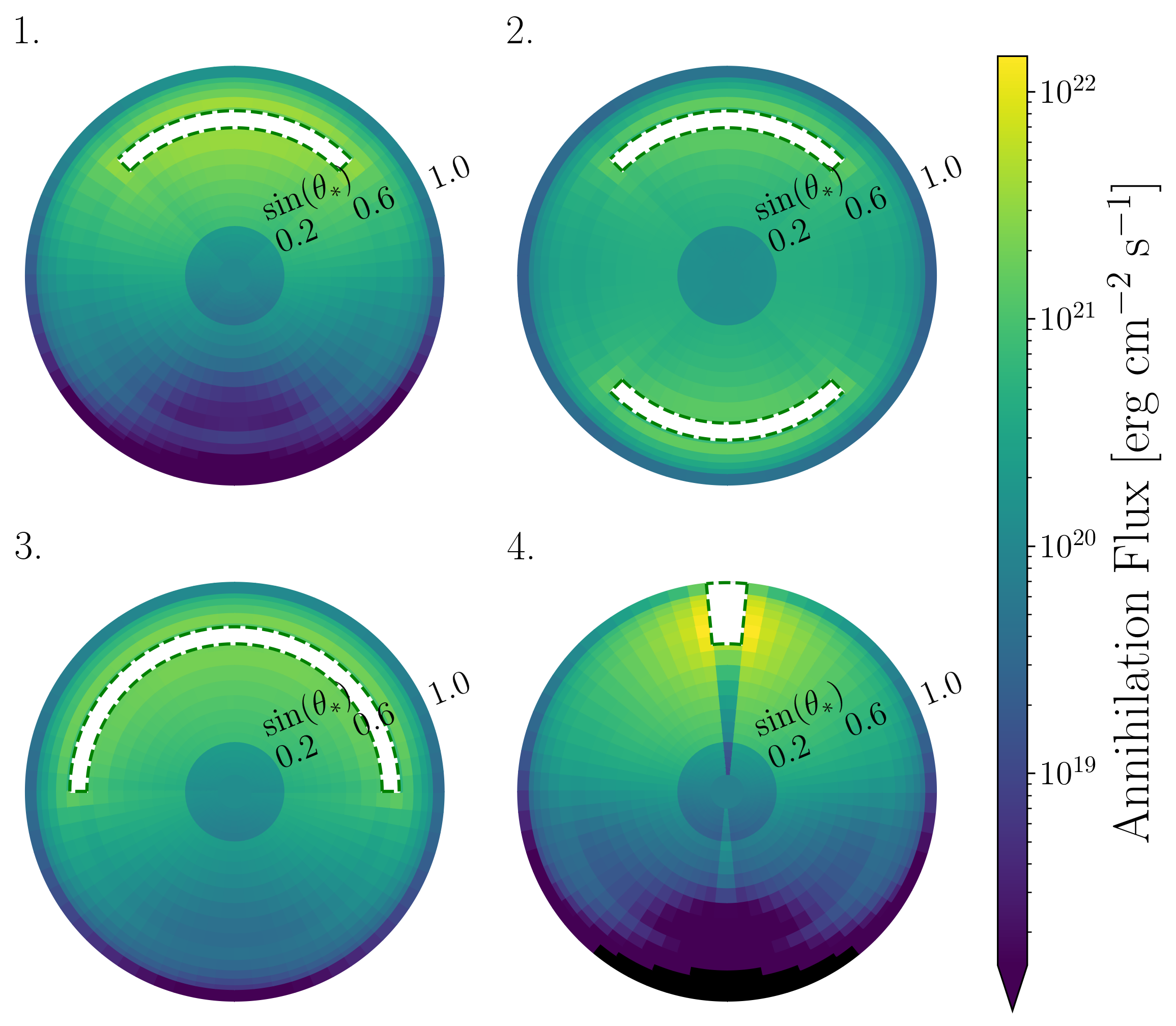} 
        \caption{Left panel:  Surface energy flux of X-rays and gamma rays hitting the star,
        multiplied by $1/2$ to estimate the effects of reflection.
        Right panel: net energy flux of electron-positron pairs absorbed and annihilating
        in the stellar atmosphere ($1/3$ of those impacting the surface; see Section 
        \ref{sec:surface_ann}).  Plots show the projection along a magnetic polar axis.  
        Fluxes are corrected for the gravitational redshift from the stellar surface to infinity.}
        \label{fig:surface_heating}
    \end{figure*}
    
    \begin{figure*}
       \centering
       \includegraphics[width = 0.48\linewidth]{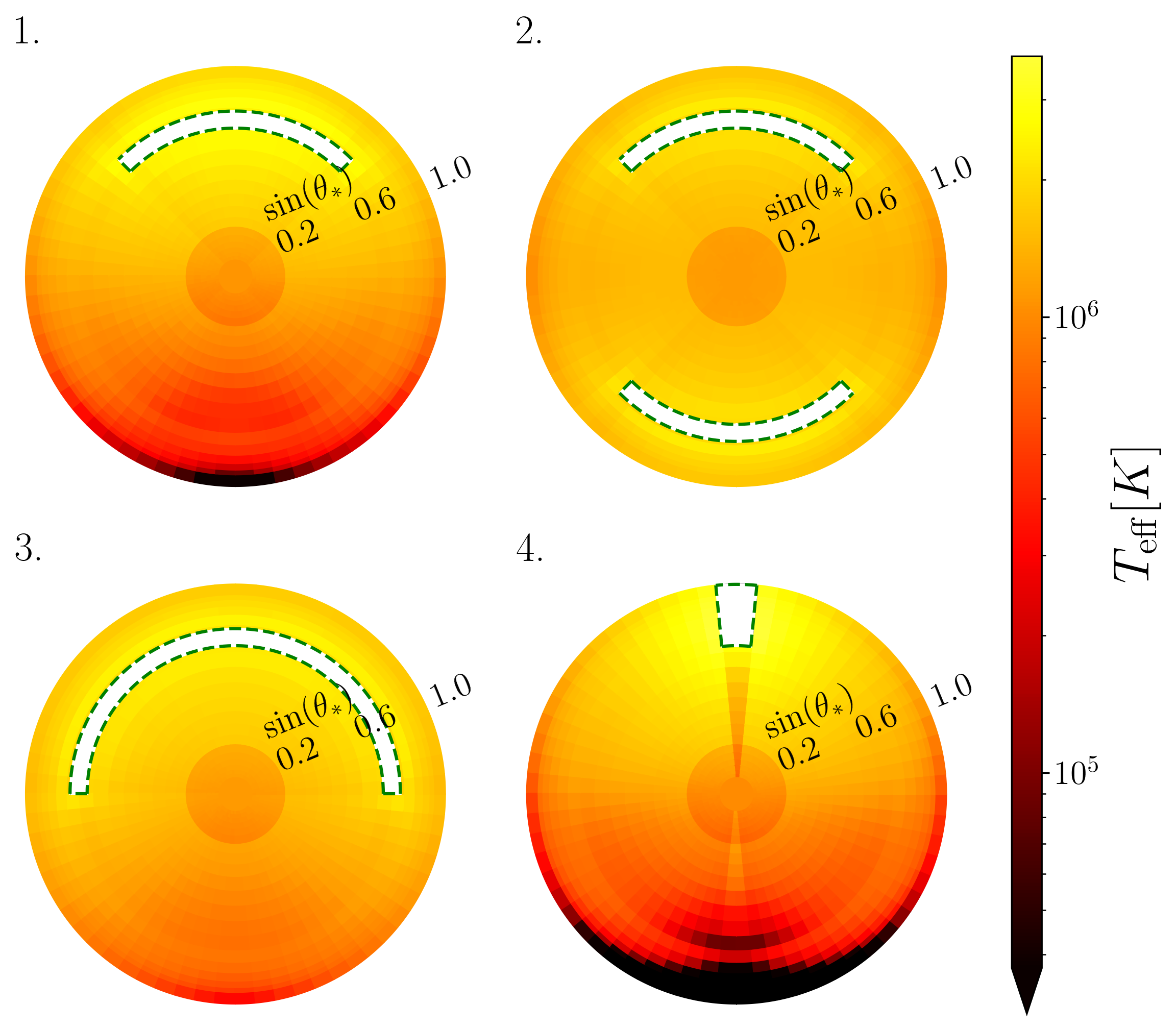}\hskip .2in
        \includegraphics[width = 0.48\linewidth]{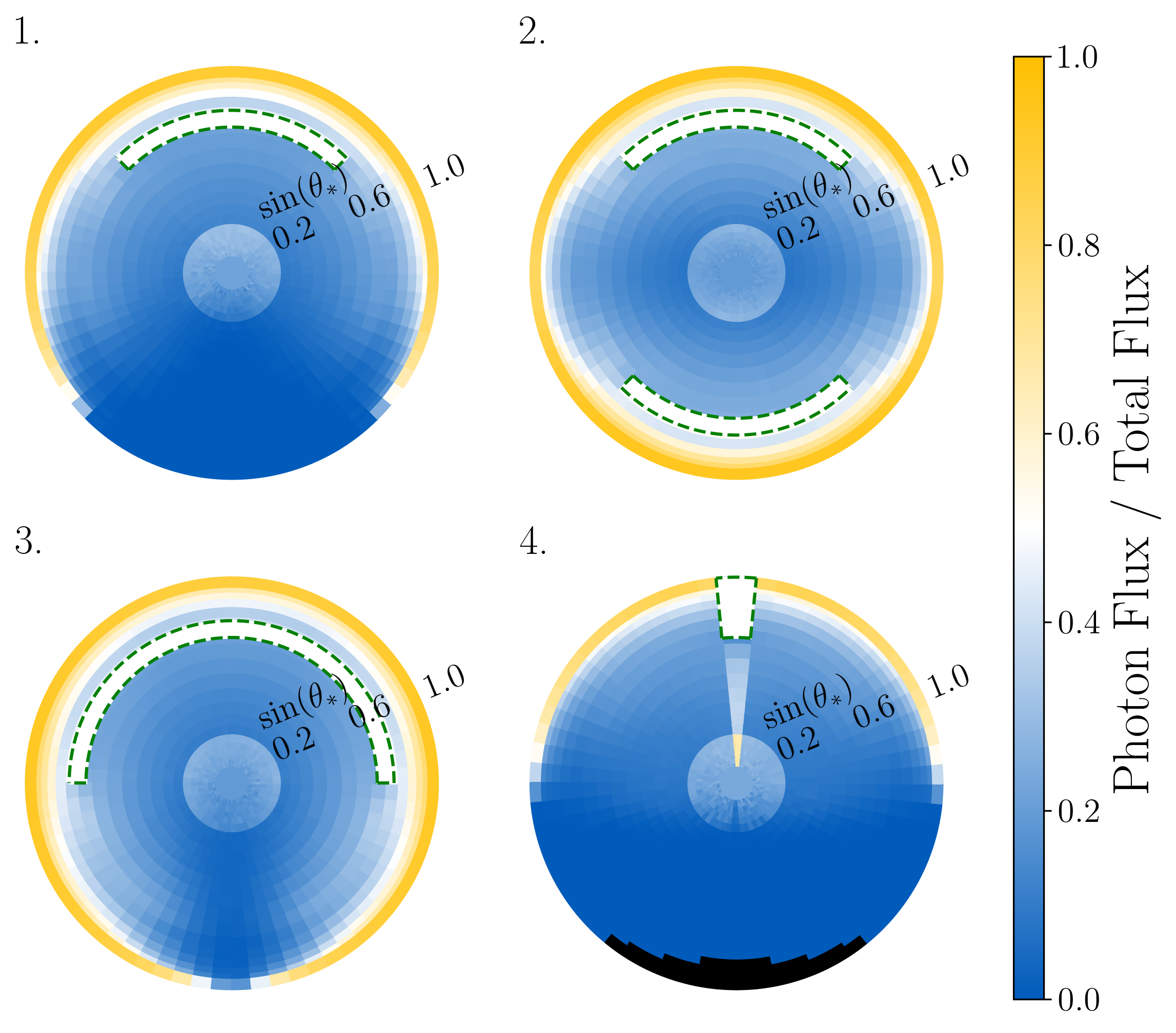}
        \caption{Left panel:  Surface effective temperature $T_{\rm eff}$ computed from
        the sum of absorbed radiation and annihilation energy fluxes in Figure \ref{fig:surface_heat}, and neglecting the contribution of
        internal heating.
        Right panel: energy absorbed from photon impact relative to total energy absorbed.
        Yellow (blue) mark parts of the surface where annihilation (radiation) heating dominates.}
    \label{fig:surface_temp}
    \end{figure*}

  \subsection{Magnetospheric Pair Density Distribution}\label{s:npair}

   We first show results for the models with polar magnetic field $10\,B_{\rm Q}$
   and higher cross section for conversion of gamma rays to pairs.  The case $B_{\rm pole} = 4\,B_{\rm Q}$ 
   is treated separately in Section \ref{sec:4BQ}.  
   
   Table \ref{tab:Temperature}
   lists the rates at which gamma ray energy is converted to pairs (column b), absorbed
   by the magnetar surface (columns c and e), and converted back to photons by volumetric 
   annihilation (column d).  The pairs impacting the surface are assumed to
   move at speed $\beta = 0.6$, which is chosen to be larger than the mean $\beta = 0.4$ of the pairs
   created by $\gamma-\gamma$ collisions (as inferred from the MC simulation) 
   to account for the additional
   energy gain from the gravitational potential.  As a result, the sum of columns (c) and
   (d) is slightly larger than (b).  The surface luminosity sums $1/3$ the pair energy flowing through the 
   stellar surface (see Section \ref{sec:surface_ann}) and $1/2$ the radiation energy flux incident
   directly on the surface;  the latter factor accounts approximately for the effects of oblique
   incidence and reflection.  The effective temperature is computed from this absorbed energy flux.  
   
   A polar view of the near-surface particle density distribution is shown in Figure \ref{fig:equil_dens}.
   Note that the pair density above the magnetar surface decreases as $\sim r^{-3}$.
   The following features will be noted.
\vskip .1in
\noindent    
    1.  When the emission structure covers less than a quadrant in azimuth, pair creation 
    is reduced in the antipodal zone. 
    This effect is strongest for the thin poloidal wedge emission structure (model 4):
    in this case, the emission is too collimated for gravitational lensing to fill in the flux
    on the opposite side of the star, leaving a strongly pair-depleted shadow.
    However, some enhancement due to lensing is seen directly antipodal to the emitting wedge.
\vskip 0.05in
\noindent
    2. The plasma density is lower near the poles, on field lines extending out to the keV resonant
    scattering radius $r_{\rm res}$ (Equation (\ref{eq:rres})), where they are pushed outward. 
\vskip .05in    
\noindent
    3. A dissipation rate $L_0 = 10^{35}$ erg s$^{-1}$ is an
   approximate upper bound for optically thin radiation transport through a magnetic field
   with polar strength $10\,B_{\rm Q}$.  Most gamma ray collisions are concentrated close to the
   emission surface around the two most localized structures (the quarter shell, model 1, and 
   thin poloidal wedge, model 4). This has the effect of pushing the emission surface outward
   from the edge of the twist zone where ohmic dissipation is concentrated.  In model 4, one half
   of the annihilations occur in the set of grid cells immediately adjacent to the wedge
   (see Table \ref{tab:structure_case}).   Note also that the characteristic O-mode X-ray scattering depth 
   above the stellar surface is reduced by a factor $0.1-0.2$ compared with the 
   quantity $\sigma_{\rm T}2n_{p,\rm NS} R_{\rm NS}$ plotted in Figure \ref{fig:equil_dens} 
   after taking into account that $n_p$ varies over a length scale $\sim R_{\rm NS}/3$, 
   the effect of ray orientation on scattering ($\sigma_{\rm O} \simeq \sin^2\theta_{kB}\sigma_{\rm T}$), 
   and the reduction in density above the magnetar surface, $n_p(r) \propto r^{-3}$.

\begin{figure*}
   \centering
    \includegraphics[width = 0.9\linewidth]{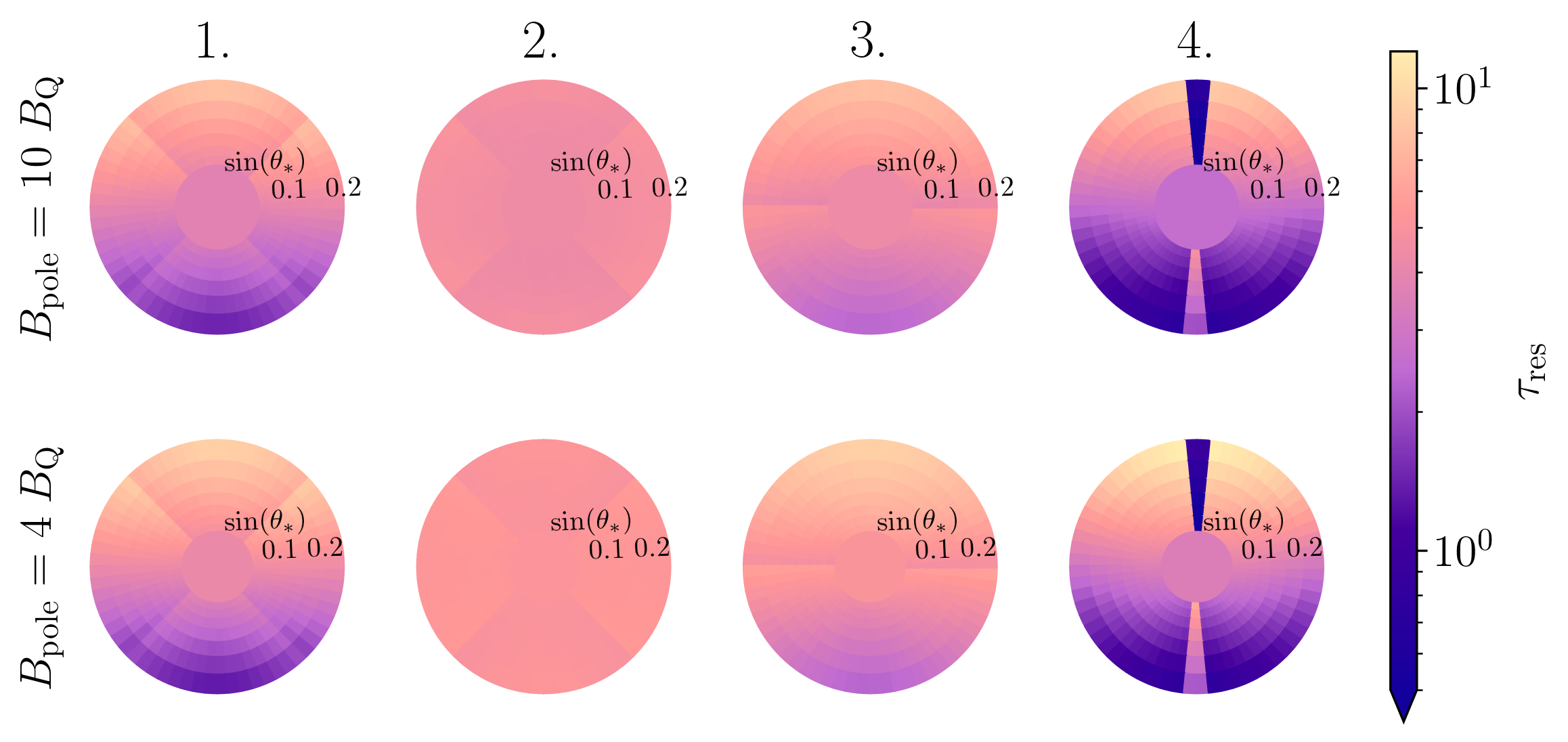}
    \caption{Optical depth $\tau_{\rm res}$ to electron cyclotron scattering, from Equation 
    (\ref{eq:optical_resonance}), plotted for each field bundle anchored near the polar cap. Results are 
    presented for (top row) $B_{\rm pole} = 10 \; B_{\rm Q}$ and $L_0 = 10^{35}$ erg s$^{-1}$, 
    and (bottom row) $B_{\rm pole} = 4 \; B_{\rm Q}$ and $L_0 = 10^{35}$ erg s$^{-1}$.
    View is the same polar projection as in Figures \ref{fig:equil_dens}$-$\ref{fig:surface_temp},
    but now focusing on the circular zone near the magnetic axis where dipole field lines
    reach the resonance radius (\ref{eq:rres}) and $n_p$ is depleted.  (For reference, the central
    pixel in this plot is now the axial grid cell.)  The physical width of the zone plotted
    grows by a factor $\sim (r_{\rm res}/R_{\rm NS})^{3/2}$ near the scattering radius $r_{\rm res}$.}
    \label{fig:optical_resonance}
\end{figure*}

    \subsection{Supporting a Weaker Global Current}

    Figure \ref{fig:twist_ratio} compares the pair density produced non-locally by $\gamma-\gamma$
    collisions with the minimum needed to support a global twist (Equation \ref{eq:twist_density}).
    In all four arcade models,
    this threshold is easily reached near the poles -- even when the global twist 
    is strong enough ($\Delta\phi_{\rm N-S} \sim 1$ radian) to induce an order-of-magnitude
    increase in spindown torque \citep{Thompson2002}.  The pair density falls below threshold only in some 
    gamma ray shadows that are antipodal to an emitting arcade.

    \subsection{Surface Heating}
    
   Figure \ref{fig:surface_heating} shows a polar view of the surface energy flux produced by 
   pair annihilation (right panel) and photon impact (left panel).  The annihilation energy
   flux is $(1/3)\cdot(1/2)\cdot 2 n_{p,\rm NS}\,\gamma\beta m_ec^3$, accounting for the proportions 
   of pairs that are absorbed rather than reflected and are moving downward to the star.  
   Direct photon impact makes a generally smaller contribution to surface heating.
      
   This surface flux is inhomogeneous and will contribute to lower-energy X-ray pulsations.  
   Note also that heating from positron annihilation is more evenly distributed across the stellar
   surface, as the magnetic 
   field can collect pairs farther out in the magnetosphere and channel them toward the surface.
   
   The corresponding effective temperature
   is shown in the left panel of Figure \ref{fig:surface_temp}, and the relative proportions of the
   two energy fluxes in the right panel. (The effective temperature is larger by a factor $2^{1/4}$
   than plotted if one allows only for re-radiation into the O-mode.)  
   
   We have
   not attempted to model the frequency distribution of the radiation re-emitted from the surface,
   comprising photons emitted directly from positron annihilation or resulting from partial thermal equilibration with the atmosphere.
   Following the discussion in Section \ref{sec:surface_ann}, most of the annihilation radiation is 
   concentrated at an O-mode scattering depth less than unity (Equation (\ref{eq:anndepth})).  
   When the magnetospheric O-mode depth is low,
   a significant part of the radiation re-emitted in the X-ray band will therefore follow the same
   annihilation bremsstrahlung spectrum as the primary gamma-ray emitting arcades.

\begin{figure*}
    \centering
    \includegraphics[width = 0.55\linewidth]{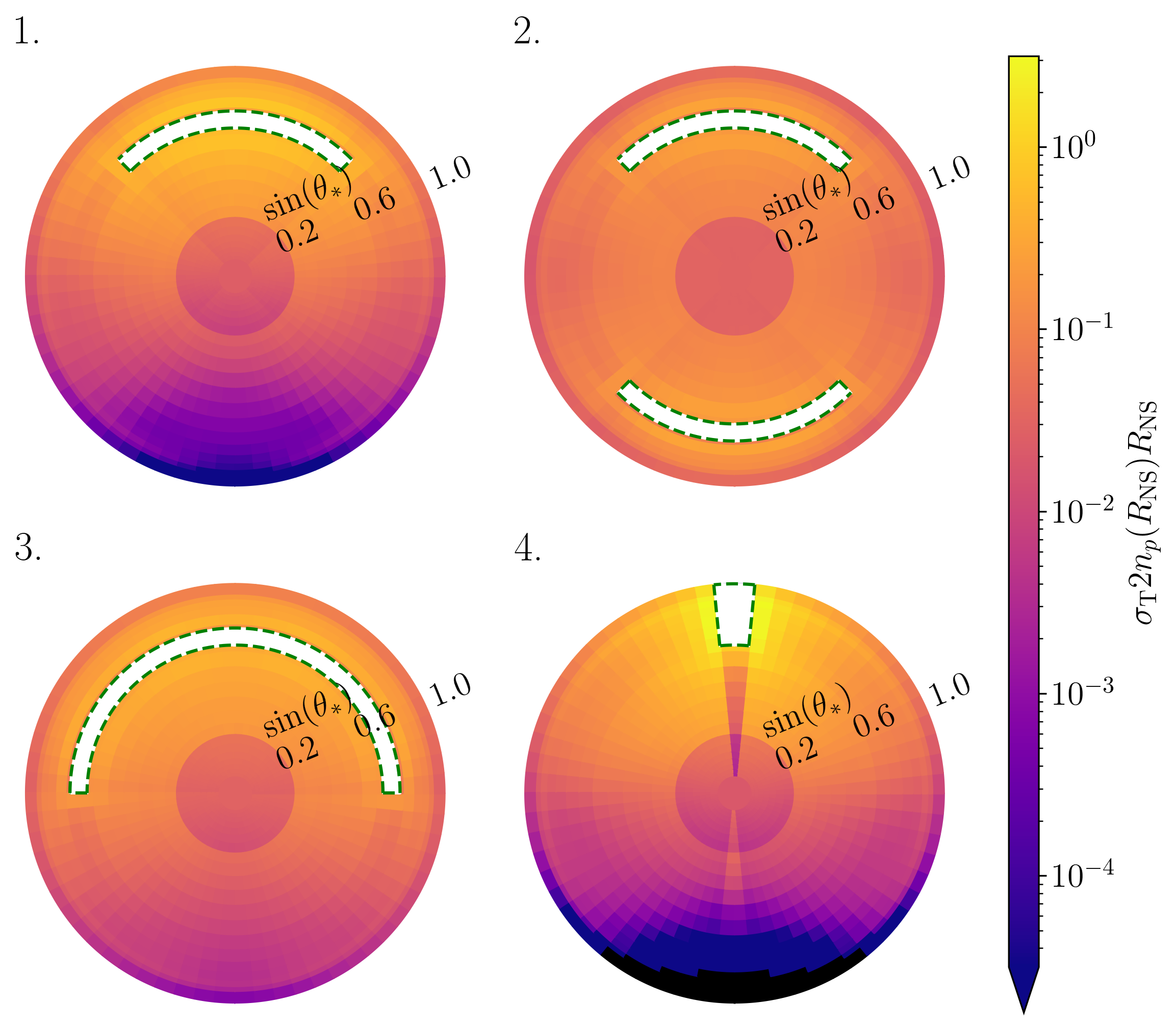}
    \caption{Equilibrium density $n_{e^+} + n_{e^-}$ at the surface of the star, as in Figure
    \ref{fig:equil_dens}, but now for a lower polar field $4\,B_{\rm Q}$ and higher luminosity
    $L_0 = 10^{35}$ erg s$^{-1}$.}
    \label{fig:surf_dens_lowB}
\end{figure*}

\begin{figure*}
    \centering
    \includegraphics[width = 0.9\linewidth]{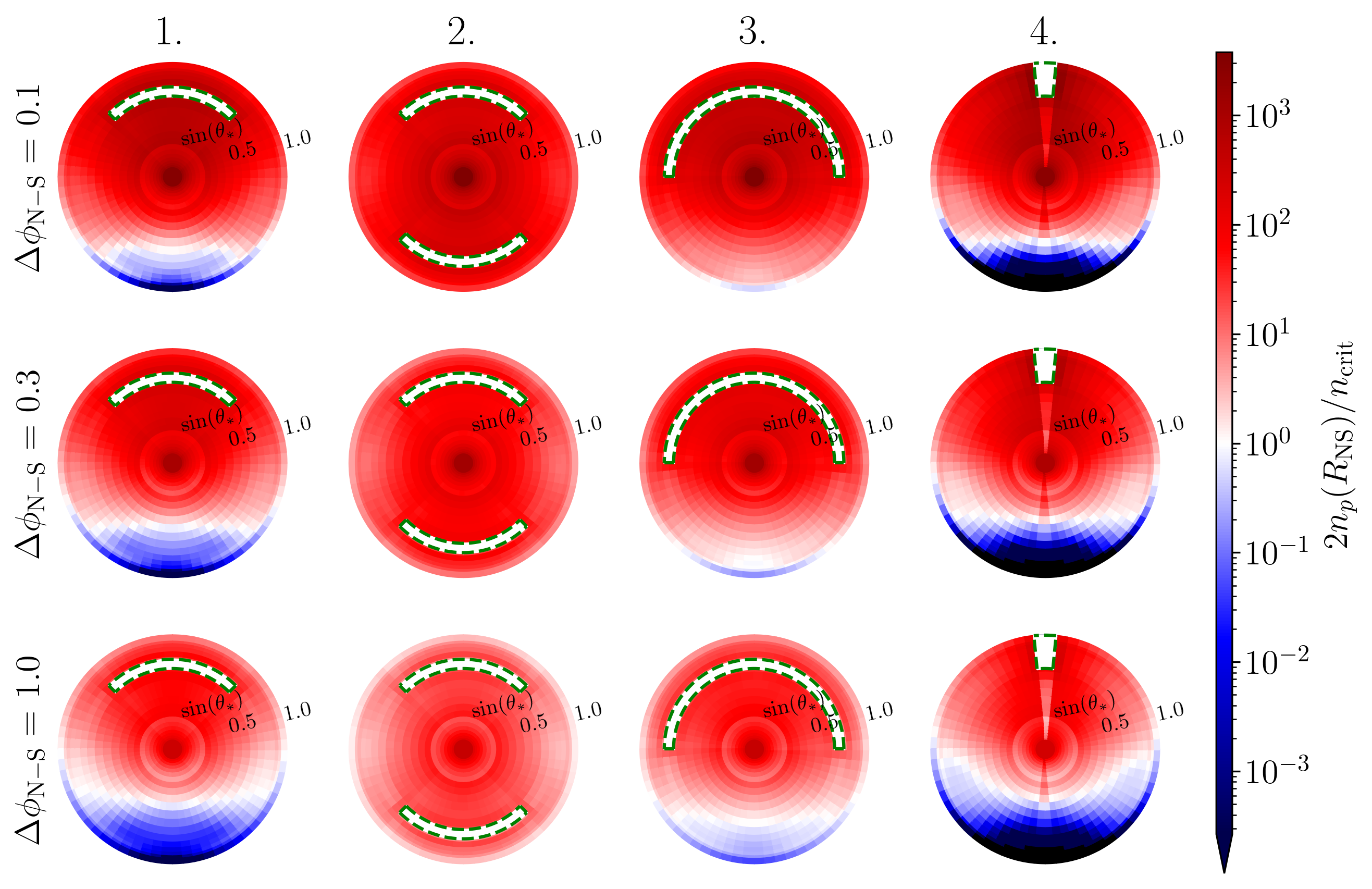}
    \caption{Comparison of the plasma density to the critical density needed to support a global 
    magnetospheric twist, as in Figure \ref{fig:twist_ratio}, but now for a lower polar 
    field $4\,B_{\rm Q}$ and luminosity $L_0 = 10^{35}$ erg s$^{-1}$.}
    \label{fig:dens_ratio_lowB}
\end{figure*}

    \begin{figure*}
        \centering
        \includegraphics[width = 0.48\linewidth]{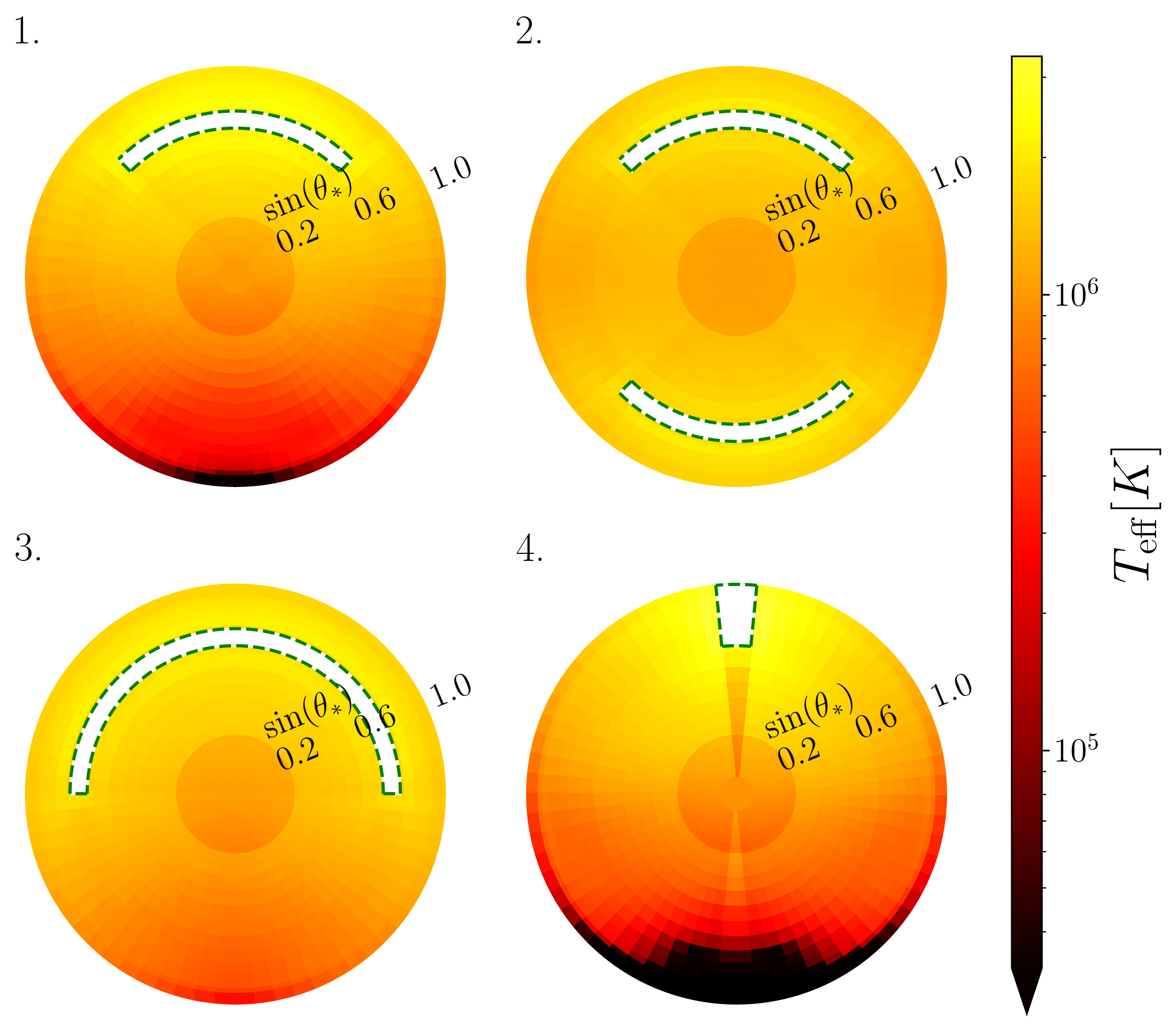}\hskip .2in
        \includegraphics[width = 0.48\linewidth] {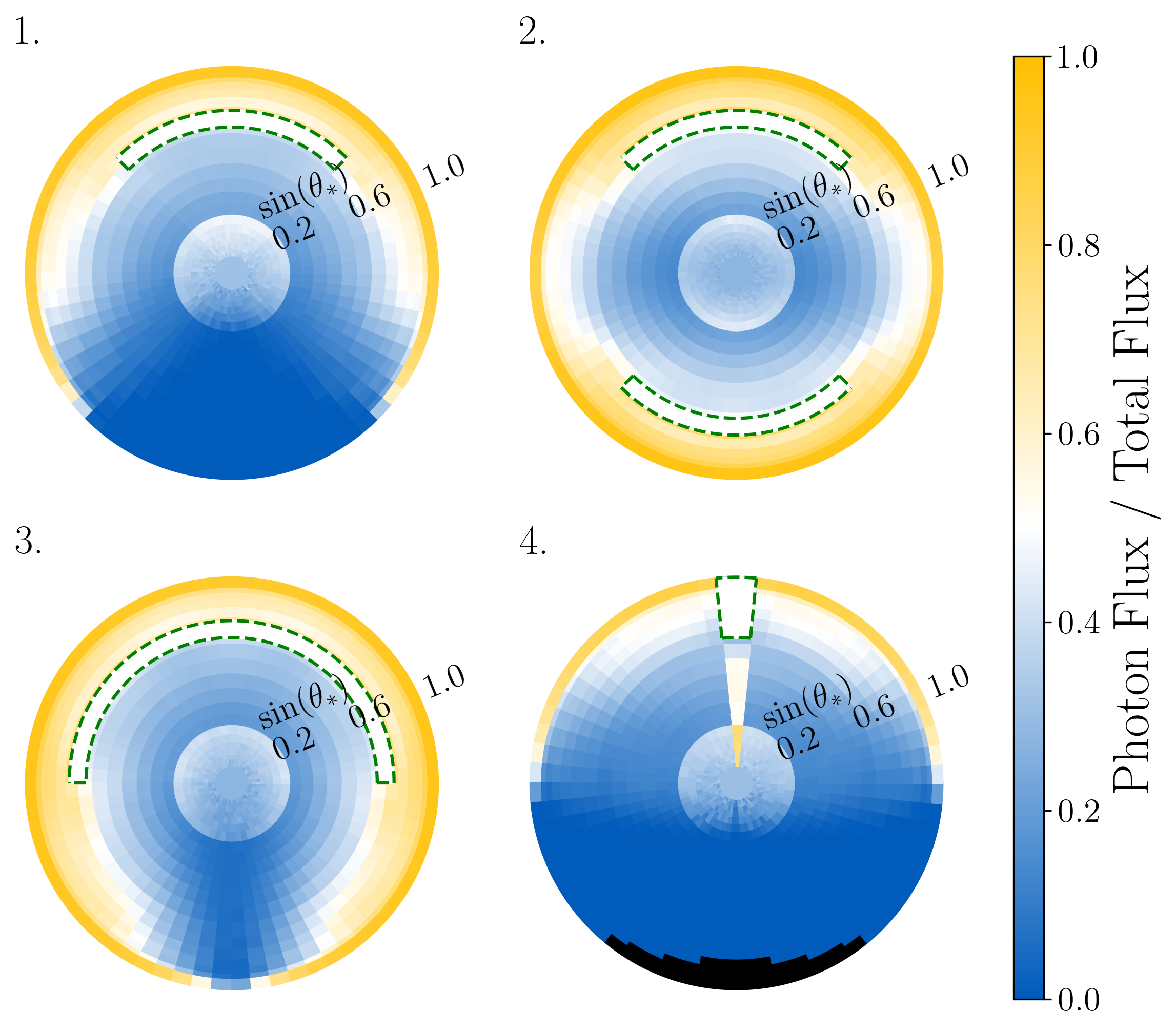} 
        \caption{Surface effective temperature (left panel) and comparison of 
        absorbed photon and annihilation heat fluxes (right panel), as in Figure \ref{fig:surface_temp},
        but now for a lower polar field $4\,B_{\rm Q}$ and luminosity $10^{35}$ erg s$^{-1}$.}
        \label{fig:surface_temp_lowb}
    \end{figure*}

\subsection{Resonant Scattering Corona}

An outflow of trans-relativistic $e^\pm$ near the pole of a magnetar will scatter
X-ray photons emitted closer to the star.  The dominant contribution to the 
scattering opacity comes from the electron cyclotron resonance.  For photons in the $1-10$ keV range,
this resonance is significantly displaced from the surface of the star (see Equation 
(\ref{eq:rres})).  To gauge the impact
of cyclotron scattering, we consider the interaction near the magnetic pole, where
photons and $e^\pm$ move nearly radially.  

The $e^\pm$ feel such a strong drag force that their
motion is tied to the angular dispersion of the photons \citep{Thompson2008a,Beloborodov2013b};
we approximate their motion
as cold with speed $\beta c$.  The position of the resonance is obtained from
Equation (\ref{eq:rres}) in terms of the Doppler-shifted photon frequency 
$\omega' = \gamma(1-\beta)\omega$ and the polar magnetic field $B_{\rm pole}$.
The optical depth is obtained by integrating over the cross section $\sigma_{\rm res} = (2\pi^2e^2/m_ec)\delta(\omega'-\omega_{ce})$, giving
\ba\label{eq:tau_res}
\tau_{\rm res} &\;=\;& \int dr\, (1-\beta)2n_p(r)\,\sigma_{\rm res}\nn
&\;=\;& {\pi\over 4\alpha_{\rm em}} {(1-\beta)^{2/3}\tau_{\rm T,NS}\over\gamma^{1/3}}
{(m_ec^2/\hbar\omega)^{1/3}\over (B_{\rm pole}/ B_{\rm Q})^{2/3}} \nn
&\;=\;& 93\,\tau_{\rm T,NS} {({\rm keV}/\hbar\omega)^{1/3}\over (B_{\rm pole}/ 10 \, B_{\rm Q})^{2/3}}.
\qquad (\beta=0.6)\nn
\label{eq:optical_resonance}
\ea
Here, we use the shorthand $\tau_{\rm T,NS} \equiv \sigma_{\rm T} \cdot 2n_{p,\rm NS}R_{\rm NS}$
and apply the radial scalings $n_p(r) = (r/R_{\rm NS})^{-3}n_{p,\rm NS}$ and $\omega_{ce} \propto r^{-3}$.

The result is plotted in the top row of Figure \ref{fig:optical_resonance}, focusing on the circular zone
surrounding the pole in Figure \ref{fig:equil_dens} where the pair density is depleted due to outflow.
For the adopted gamma ray luminosity of $10^{35}$ erg s$^{-1}$, the optical depth reaches
$\tau_{\rm res} \sim 5-10$ along much of this field bundle.  This shows that $L_0 \sim 10^{35}$
erg s$^{-1}$ is an approximate lower bound to the gamma ray luminosity of a magnetar
that seeds enough trans-relativistic pairs to 
significantly rescatter X-rays around the keV blackbody peak.

The pair density is distributed inhomogeneously across this extended magnetic field bundle.
This inhomogeneous distribution will translate into azimuthal variations in $\tau_{\rm res}$
within the resonant scattering zone.  In this way, we obtain a direct connection between
azimuthal variation in the yielding profile of the magnetar crust at large distances from the
poles, which translate into azimuthal variations in the flux of annihilation radiation in 
the inner magnetosphere, and thence into measurable changes in the pulse profile in the lower-frequency
X-ray band.

\subsection{Lower Polar Field Strength:  1E 2259$+$586}\label{sec:4BQ}

When the polar field is reduced from $10\,B_{\rm Q}$ to $4\,B_{\rm Q}$, while fixing $L_0$,
the efficiency of pair creation is reduced by a factor $\sim 2$.
This is mainly due
to the reduction in the cross section for $\gamma-\gamma$ collisions ($\sigma_{\rm \gamma\gamma}
\propto B/B_{\rm Q}$ when $B \gtrsim B_{\rm Q}$), and also due to the tighter kinematic
constraints on two-photon pair creation in the weaker field.

Figures \ref{fig:surf_dens_lowB}$-$\ref{fig:surface_temp_lowb} show the pair density,
the ratio of the pair density to the critical density supporting a global twist, and
the surface energy flux and temperature for all four arcade models.   Comparing with 
Figures \ref{fig:equil_dens}, \ref{fig:twist_ratio} and \ref{fig:surface_temp} for the
case $B_{\rm pole} = 10\,B_{\rm Q}$,
one sees a modest expansion in the size of the
zone where $n_p$ is smaller than the critical density that will support a global twist.
This effect is most prominent in the gamma ray shadow opposite the emitting arcades,
and is partly compensated by a reduction in the current
density associated with a given global twist angle $\Delta\phi_{\rm N-S}$.

The decreases in pair density and in $B_{\rm pole}$ have opposing effects on the optical depth 
to electron cyclotron scattering (Equation (\ref{eq:tau_res})).
As a result we find $\tau_{\rm res} \sim 5-10$
on much of the polar field bundle reaching radius $r_{\rm res} \sim 15\, R_{\rm NS}$;  see the
bottom row of Figure \ref{fig:optical_resonance}.

\section{Implications for Magnetar Behavior}\label{sec:summary}

The problem posed by magnetar electrodynamics has a few aspects.  1. Where do currents flow
outside the star and how smooth is the pattern of crustal shearing?
2. How is $e^\pm$ plasma sustained in the magnetosphere and in what kinetic state?  
3. How closely tied is the energetically dominant hard X-ray emission of a quiescent magnetar 
to the processes sustaining the current?  

A popular approach to modelling the hard X-ray emission of a magnetar is to
posit the acceleration of relativistic charges in parts of the magnetosphere where $B < B_{\rm Q}$.
Outflowing charges reaching $150-300$ km from the star will resonantly upscatter keV 
photons with a hard spectrum \citep{Fernandez2007,Baring2007}.  This process has been
connected self-consistently to the relativistic double layer plasma state \citep{Beloborodov2013a,Wadiasingh2018}.
It is challenging in this picture to explain a rapid increase in hard X-ray output that is
{\it not} accompanied by a prompt surge in the spindown torque.  
The rapidity of the luminosity increase implies a 
concentration of the driving crustal motions around the pole.

\cite{Thompson2020} demonstrated how hard-spectrum X-ray emission follows
directly from a state of pair annihilation-creation equilibrium in zones of high current
density.  These twist zones could form anywhere in the inner magnetosphere. 
The emission spectrum produced at moderate dissipation rates
carries the direct signature of a QED process ($e^\pm$ annihilation) operating in magnetic fields above
$\sim 10^{14}$ G.  We have examined the global consequences of this
process, especially for the polar zones that control the spindown torque, radio emission
and emergent X-ray pulse profile.

With this aim, we developed a MC description of non-local pair creation around a magnetar.   The rate for
$\gamma + \gamma \rightarrow e^+ + e^-$ is enhanced by a factor $\sim B/B_{\rm Q}$
in the super-strong magnetic field;  the kinematic constraints on $e^\pm$ pair creation are
also relaxed compared with an unmagnetized vacuum. 
We consider a range of dissipative structures anchored at intermediate latitudes,
away from the magnetic poles, and include the expansion in gamma ray exposure due to gravitational lensing.
These calculations provide a benchmark demonstrating the sort of resolution required to represent the
physics in a fully three-dimensional and general relativistic particle-in-cell simulation.

Our calculations apply to a source like 1E 2259$+$586 with a hard X-ray continuum
of moderate intensity, and a lower optical depth to gamma ray collisions than the
brightest persistent magnetars.
Even when the measured $15-60$ keV flux is around $10^{34}$ erg s$^{-1}$, corresponding
to a bolometric gamma ray flux near $10^{35}$ erg s$^{-1}$, the pair creation
rate in the polar zones is easily high enough to support a magnetic twist
$\Delta\phi_{\rm N-S} \sim 0.1-1$ radians.  Enough particles are generated to short
out the relativistic double-layer structure that would form if the magnetosphere were more smoothly 
twisted \citep{BT2007}.

More generally, the quiescent magnetar 1E 2259$+$586 can be viewed as
a Rosetta Stone for understanding the electrodynamics of magnetars and their broadband
electromagnetic emission.  Its hard $15-60$ keV spectrum is
consistent with pair annihilation in a magnetic field around $B_{\rm Q}$
(see Figure 3 of \citealt{Thompson2020}).  By contrast, the spin history of
1E 2259$+$586 is consistent with weak twisting of the polar magnetic field.
Its overall behavior fits well with our inference that hard X-ray emission and associated
optical-IR plasma emission originate in the inner magnetosphere.

We next summarize some implications of the MC results presented here.

\subsection {Predictions and Implications for Magnetar Physics}

1. {\it Gamma ray spectrum.}  The X-ray continuum measured up to $50-100$ keV is predicted to extend up 
to $\sim 0.5-1$ MeV, but then cut off sharply.   (See the sample spectra in Figure 14 of \citealt{Thompson2020}.)
This continuum bears a direct imprint of QED processes operating 
in super-Schwinger magnetic fields.  In magnetic fields stronger than $B_{\rm Q}$, 
the two-photon decay of an electron-positron pair becomes
a soft-photon correction to single-photon decay, with a flat photon spectrum.

2. {\it Pseudofaults.}  Zones of concentrated magnetospheric shear, which host collisional pair plasma,
are connected to narrow shear zones of a sub-kilometer width in the solid crust.
Magnetar outbursts provide strong evidence for such structures.  The durations
of most short X-ray bursts are comparable to the time for solid stresses to adjust
globally in the crust \citep{Gogus2001}, even while high-temperature post-burst afterglow is 
concentrated in a tiny fraction of the stellar surface (e.g. \citealt{Woods2004}).  These structures have also
been shown to arise in a global elastic-plastic-thermal model of the magnetar crust
\citep{Thompson2017}.

3. {\it X-ray reprocessing by cyclotron scattering.}
Non-local $e^\pm$ creation in the polar zones has significant implications for the emergent X-ray
polarization and X-ray pulse profiles.    The flux of trans-relativistic 
pairs moving out beyond $\sim 15-30$ stellar radii is high enough to
generate an optical depth $6-40$ to electron cyclotron scattering 
(Figure \ref{fig:optical_resonance}),  even for a gamma ray luminosity $L_0 = 10^{35}$ erg s$^{-1}$
adjusted to the modest nonthermal output of 1E 2259$+$586.   The inhomogeneity of the pair creation 
profile above the polar cap translates into axial variations in the optical 
depth to resonant scattering, which are a source of pulsations.

At the same time, the pairs that resonantly scatter keV photons
will add negligibly to the energy of the scattered X-rays.   The kinetic energy
carried by the pairs within a polar angle $\theta_\star \sim (r_{\rm res}/R_{\rm NS})^{-1/2}
= 0.2\,(B_{\rm pole}/10\,B_{\rm Q})^{-1/6}(\hbar\omega/{\rm keV})^{1/6}$ is
\ba
L_{\rm kin} &\;=\;& \pi\theta_\star^2 R_{\rm NS}^2\cdot 2n_{p,\rm NS} (\gamma-1)\beta m_ec^3\nn
        &\;\sim\;& 9\times 10^{31}\,\left({\tau_{\rm res}\over 10}\right)
        \left({\hbar\omega\over {\rm keV}}\right)^{2/3}
        \left({B_{\rm pole}\over 10\,B_{\rm Q}}\right)^{1/3}
\; {\rm erg~s^{-1}}.\nn
\ea

A background polar magnetic twist $\Delta\phi_{\rm N-S}$ will enforce a small
differential drift of the $e^+$ and $e^-$, with a speed $\Delta\beta_{+-} \sim n_{\rm crit}/n_p$, 
where $n_{\rm crit} = J/2e\beta c$ is given by Equation (\ref{eq:twist_density}).
This differential drift causes an average frequency shift $\sim \Delta\beta_{+-}^2$ per scattering,
thereby driving a fractional increase $\sim \tau_{\rm res}\Delta\beta_{+-}^2 \propto \tau_{\rm res}^{-1}$ 
in the X-ray power.  This change is modest even for large twist, given
that $\Delta\beta_{+-} \sim \Delta\phi_{\rm N-S}/\tau_{\rm res}$ in the resonant scattering zone.

As a result, the trans-relativistic pairs
are expected to have a weaker effect on the emergent X-ray spectrum than in the calculations of
\cite{Fernandez2007} and \cite{Nobili2008}, where the particle density was tied to the critical density
needed to support the local magnetic twist, and in the model of \cite{Beloborodov2013a,Wadiasingh2018}, where
the pairs are injected relativistically above the polar cap and are
 the main
energy source for the hard X-ray continuum.  

4.  {\it Emergent X-ray polarization.}
A significant flux of E-mode photons will be generated by scattering from the O-mode.
(The gyrating electron couples to both polarization modes, the coupling to the
O-mode being suppressed by a factor $(\cos\theta_{kB}^O)^2 = (\hat k^O\cdot\hat B)^2$ 
compared with the coupling to the E-mode.)   In fact, at the high optical depths
considered, a simple argument suggests that the net polarization is significantly reduced.
In contrast with the case of $e^\pm$ drift at low optical depth
in a fixed central radiation field \citep{Beloborodov2013b}, 
the photon flow must come into approximate alignment with the magnetic field
within the resonance.  We balance the rates
of scattering from O to E and E to O and make use of the cross sections \citep{Harding2006}
$d\sigma_{\rm res}^{O\rightarrow E}/d\Omega_E
= (\cos\theta_{kB}^O)^2 (3\pi^2 r_ec/8\pi) \delta(\omega'-\omega_{ce})$ and
$d\sigma_{\rm res}^{E\rightarrow O}/d\Omega_O = (\cos\theta_{kB}^O)^2(3\pi^2 r_ec/8\pi) \delta(\omega'-\omega_{ce})$.  
This gives $n_E = n_O$ in each angular direction at high optical depth with multiple scattering. Photons of both
polarization modes scatter at the same rate when propagating along ${\bm B}$.
There is an interesting potential connection
here with measurements of $10-30\%$ net polarization in the 2-4 keV band
by IXPE in the three magnetars 4U 0142$+$61, 1E 2259$+$586, and 1E 1841$-$045 \citep{Taverna2022,Heyl2024,Rigoselli2024}.

5. {\it Surface composition.}
The energies of the particles hitting the magnetar surface (both pairs and gamma rays) are
cut off sharply above $\sim 1$ MeV. Knockout of neutrons and protons out of
of heavy nuclei requires higher energies of at least 15-20 MeV \citep{Dietrich1988}.  As a result, a heavy
atmospheric composition and a condensed state \citep{Lai2001} are more plausibly maintained when the plasma
around the magnetar is collisional and trans-relativistic, than when it is composed of
relativistic $e^\pm$ streaming with energies of hundreds of MeV \citep{BT2007}.

6. {\it Correlation between hard X-ray and IR emission.}
The enhanced IR-optical emission of magnetars is naturally connected with the collisional
zones of hard X-ray emitting plasma, where the plasma frequency lies in this band
\citep{Thompson2020}.  Significant correlations between changes in hard X-ray and IR
output should therefore be present; there is already evidence for such a correlation in softer X-ray
bands following the 2003 outburst of 1E 2259$+$586 \citep{Tam2004}.   The rapid and persistent 
changes that are detected in the magnetar radio pulse profile (e.g. \citealt{Huang2023}) are 
consistent with the radio emission being concentrated much closer 
to the separatrix between open and closed magnetic field lines \citep{Thompson2008b}.

\subsection{Higher Luminosity and Stronger Magnetic Field}

Some magnetars emit $15-60$ keV radiation with a luminosity 100 times higher
than 1E 2259$+$586:  these include the most active burst sources SGR 1806$-$20 and SGR 1900$+$14 
and also the less active anomalous X-ray pulsar 1E 1841$-$045 \citep{Enoto2017}. 
The $15-60$ keV X-ray spectra are slightly softer ($\Gamma_X \simeq -1$)
in the latter two sources, consistent with annihilation bremsstrahlung 
in magnetic fields stronger than $4\,B_{\rm Q}$.
Indeed, the polar magnetic fields inferred from spindown are greater than
$10^{15}$ G in all three sources.  Much of the stellar surface may be threaded by fields approaching
$100\,B_{\rm Q} \sim 5\times 10^{15}$ G.  (The case for high magnetization is strongest
for SGR 1806$-$20, which emitted a gamma ray flare
in 2004 carrying an energy $\sim 5\times 10^{46}$ erg, approaching the total energy in a
$10^{15}$ G dipole magnetic field; \citealt{Hurley2005,Terasawa2005}.)

In these luminous and strongly magnetized 
sources, the optical depth to $\gamma-\gamma$ collisions is high enough that
most gamma-rays of energy $\gtrsim m_ec^2$ 
emitted from the twist zones will be absorbed over a fraction of the stellar
radius.  (A hint of this effect is seen in our narrow wedge model 4 with 
$B_{\rm pole}=10\,B_{\rm Q}$; see Figure \ref{fig:equil_dens}.)  The most energetic
annihilation bremsstrahlung photons (with frequencies just below the threshold 
$\omega_{1\gamma}$ for direct pair conversion) experience an additional resonant enhancement in their cross section for electron scattering.\footnote{This resonance
involves a vertex between the incoming electron, a virtual positron, and the final-state
photon \citep{Kostenko2018}.}  In this scattering channel, the final-state photon
is usually above threshold for immediate conversion to a pair, yielding
$\gamma + e^\pm \rightarrow e^+ + e^- + e^\pm$ \citep{Kostenko2018}.  Incorporating
this effect accurately in a MC calculation requires angle-dependent frequency binning,
which is challenging in a nonuniform  magnetic field.

The strong magnetic field acts as a capacious reservoir for pairs, with the rate of two-photon 
annihilation being suppressed by a factor $(B/B_{\rm Q})^{-1}$.  
In annihilation-creation equilibrium, the ratio of pair
and photon densities is given by Equation (\ref{eq:nratio}).  The rate of emission of
gamma rays with energy $> m_ec^2$ 
in our adopted spectrum (\ref{eq:emspec}) is $0.24 L_0/m_ec^2$.  Balancing
isotropic emission from arcade structures with area $A$ with annihilation in nearby magnetic
field, one obtains an equilibrium pair density $\sigma_{\rm T} n_p R_{\rm NS} = 
100\,B_{15}(L_0/4\times 10^{36}~{\rm erg~s^{-1}})(A/4\pi R_{\rm NS}^2)^{-1}$.   
A similar result is obtained by balancing with surface annihilation, which is also limited
by outward radiation pressure on the downward moving pairs.  
Given such a high pair density, the flux of photons with energies
above $m_ec^2$ is rapidly degraded by electron recoil and pair conversion.
The density and temperature of the pairs must therefore drop rapidly with distance 
from the surface of the current-carrying twist zone.   

It should be emphasized that the flat bremsstrahlung number spectrum is preserved by
multiple Compton scattering by warm electrons:  the same spectral slope $dn_\gamma/d\omega
\propto \omega^{-1}$ corresponds to a uniform flux of photons upward 
in frequency space \citep{Kompaneets1957}.   When the pairs are
continuously heated (as in a zone of high current density) the formation of a
Wien peak is avoided because photons upscattered to an energy $\sim m_ec^2$ are
efficiently converted back to pairs, with a large ratio $n_p/n_\gamma(\hbar\omega > m_ec^2)$. 
A full test of the emitted X-ray spectrum
requires a modification of the leaky box MC simulation of \cite{Thompson2020} to 
include (i) inhomogeneous currents and magnetic fields; (ii) collisions between
$e^\pm$ mediated by one-photon annihilation; and (iii) self-consistent solution of
the Maxwell equation to obtain the resistive electric field.  A parallel effort
is required to model the radiative output of positron annihilation in the magnetar atmosphere, including backscattering of warm positrons and electrons (both
by 1-photon annihlation and ion collision) bremsstrahlung photon emission,
and multiple electron-photon scattering \citep{Kostenko2020}.

\vskip .2in
\noindent
This research was supported by NSERC grant RGPIN-2023-04612 and by the Simons Foundation
grant MP-SCMPS-00001470-08.

\vfil\eject
\appendix 

\section{Distribution of Pair Creation with Photon Energy}\label{s:appendix}

\begin{figure*}
    \centering
    \includegraphics[width = 0.9\linewidth]{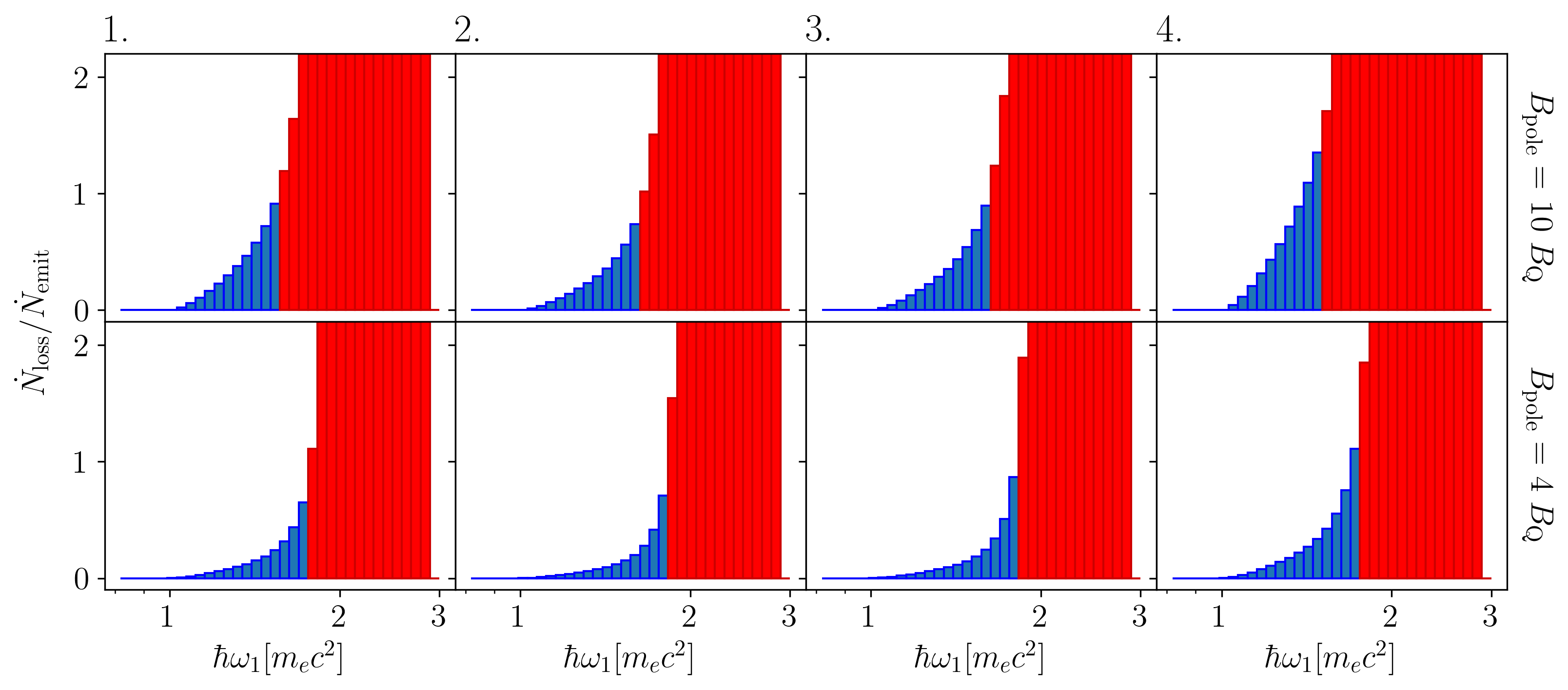}
    \caption{Ratio of the rate of photon consumption to photon emission
      rate, as binned by the energy of the more energetic photon in each collision event.
      Results are presented for (top row) $B_{\rm pole} = 10 \; B_{\rm Q}$ and (bottom row)
      $B_{\rm pole} = 4 \; B_{\rm Q}$, both with $L_0 = 10^{35}$ erg s$^{-1}$. We choose a cutoff on the upper 
    photon energy such that this ratio does not exceed $1$ for arcade
    models 1-3.  As result, the red bins do not
    contribute to the tabulated pair creation rate, corresponding to all
    photons of energy $\gtrsim (1.5-1.6)\,m_ec^2$.  Note that the adopted
    emission energy spectrum $dL_\gamma/d\ln\omega$ peaks at $\hbar\omega \simeq m_ec^2$ and is smaller by a factor 0.1 at $2m_ec^2$.  In the case of 
    arcade model 4, almost half of the pair creation is localized in the set of cells
    next to the emitting arcade, indicating an expansion of the emission surface of the pair-creating gamma rays (see Table \ref{tab:Temperature}).  We adopt a limiting ratio 1.5 for model 4.}
    \label{fig:luminosity_ratio}
\end{figure*}

Here we tabulate how $e^\pm$ pair creation is distributed with
respect to the energy carried by the more energetic photon involved in each
photon collision.  
Figure \ref{fig:luminosity_ratio} shows this quantity as normalized by 
the emission rate of the more energetic seed photon.
We excise the contribution to $\dot N_{\gamma\gamma\rightarrow e^+e^-}$
when this ratio is larger than unity in arcade models 1-3, so
as to avoid overcounting.

Photons of energy $\gtrsim 2m_ec^2$ can convert to a pair by colliding with a wide
range of soft photons, with a cross-section in the strong magnetic field 
scaling with the soft photon energy as $\omega^{-3}$ (Equation (\ref{eq:sigmagg}).
 For the MC runs with $B_{\rm pole} = 10 \; B_{\rm Q}$, in the top row of Figure \ref{fig:luminosity_ratio}, the top 14-17 energy bins are removed, lowering the overall pair production rate by $\sim 90\%$. For the models with $B_{\rm pole} = 4 \; B_{\rm Q}$, in the bottom row of Figure \ref{fig:luminosity_ratio}, only the top 13 energy bins are removed, reducing the pair production rate by $\sim 80\%$.
Photons of energy $\gtrsim 2m_ec^2$ carry less than
10 percent of the annihilation energy; they convert rapidly enough to pairs
that their energy is effectively transferred to lower energy photons by
repeated annihilations.  

In model 4, we find that almost one half of the pair creation is concentrated
in the set of cells adjacent to the arcade (Table \ref{tab:Temperature}).
The energy deposited in these pairs is mainly re-released by 
magnetospheric annihilation, thereby
expanding the effective emitting surface and sustaining the rate of pair
creation in more distant cells.

\end{document}